\documentclass[12pt]{article}

\usepackage[table]{xcolor}
\usepackage[normalem]{ulem}
\usepackage{longtable}
\usepackage{footmisc}

\usepackage{scicite}
\usepackage{ref_macros_sci}

\usepackage{comment}

\usepackage{cancel}
\usepackage[normalem]{ulem}
\usepackage{graphicx}
\usepackage{setspace}

\usepackage{amsmath}
\usepackage{amssymb}
\usepackage{xspace}
\usepackage{multirow}
\usepackage{array}
\newcolumntype{C}[1]{>{\centering\arraybackslash}p{#1}}
\usepackage{rotating}

\usepackage{siunitx}
\usepackage[flushleft]{threeparttable}

\newcommand\nicer{\textit{NICER}\xspace}
\newcommand\hst{\textit{HST}\xspace}

\newcommand{\target}{AT~2022cmc\xspace}
\newcommand{\swift}{\textit{Swift}\xspace}

\renewcommand{\figurename}{Figure}
\renewcommand{\tablename}{Table}
\usepackage[final]{pdfpages}
\usepackage{times}
\usepackage[hypcap=false]{caption} 
\usepackage{subcaption, booktabs}

\usepackage{floatrow}
\DeclareFloatVCode{somespace}{\vspace{1.667\baselineskip}}
\usepackage{hyperref}
\newenvironment{sciabstract}{%
\begin{quote} \bf}
{\end{quote}}

\newcounter{lastnote}
\newenvironment{scilastnote}{%
\setcounter{lastnote}{\value{enumiv}}%
\addtocounter{lastnote}{+1}%
\begin{list}%
{\arabic{lastnote}.}
{\setlength{\leftmargin}{.22in}}
{\setlength{\labelsep}{.5em}}}
{\end{list}}

\title{The Birth of a Relativistic Jet Following the Disruption of a Star by a Cosmological Black Hole}

\author{
Dheeraj R. Pasham$^{1}$, 
Matteo Lucchini$^{1}$, 
Tanmoy Laskar$^{2}$, \\
Benjamin P. Gompertz$^{3,4}$, 
Shubham Srivastav$^{5}$, 
Matt Nicholl$^{3,4}$,\\
Stephen J. Smartt$^{5}$, 
James C.\ A. Miller-Jones$^{6}$, 
Kate D.  Alexander$^{7}$,\\ 
Rob Fender$^{8}$, 
Graham P.\ Smith$^{4}$, 
Michael D. Fulton$^{5}$, 
Gulab Dewangan$^{9}$,
Keith Gendreau$^{10}$, \\
Eric R.~Coughlin$^{11}$,
Lauren Rhodes$^{8}$, 
Assaf Horesh$^{12}$, 
Sjoert van Velzen$^{13}$, 
Itai Sfaradi$^{12}$,\\
Muryel Guolo$^{14}$,
N. Castro~Segura$^{15}$, 
Aysha Aamer$^{3,4}$, 
Joseph P. Anderson$^{16}$,\\
Iair Arcavi$^{17,18}$, 
Seán J. Brennan$^{19}$, 
Kenneth Chambers$^{20}$, 
Panos Charalampopoulos$^{21}$,\\ 
Ting-Wan Chen$^{22}$, 
A. Clocchiatti$^{23,24}$,
Thomas de Boer$^{20}$, 
Michel Dennefeld$^{25}$,\\
Elizabeth Ferrara$^{10}$, 
Lluís Galbany$^{26, 27}$,
Hua Gao$^{20}$,
James H. Gillanders$^{5}$,\\ 
Adelle Goodwin$^{6}$, 
Mariusz Gromadzki$^{28}$,
M Huber$^{20}$,   
Peter G.~Jonker$^{2,41}$, \\
Manasvita Joshi$^{29}$, 
Erin Kara$^{1}$, 
Thomas L. Killestein$^{30}$, 
Peter Kosec$^{1}$, \\
Daniel Kocevski$^{31}$,
Giorgos Leloudas$^{21}$,
Chien-Cheng Lin$^{20}$, 
Raffaella Margutti$^{32}$, \\
Seppo Mattila$^{33}$, 
Thomas Moore$^{5}$,
Tom\'as M\"uller-Bravo$^{26,27}$, 
Chow-Choong Ngeow$^{34}$, \\
Samantha Oates$^{3,4}$, 
Francesca Onori$^{35}$, 
Yen-Chen Pan$^{34}$,
Miguel P\'erez-Torres$^{36,37,38}$, \\
Priyanka Rani$^{9}$, 
Ronald Remillard$^{1}$,
Evan J. Ridley$^{3,4}$, 
Steve Schulze$^{42}$,\\
Xinyue Sheng$^{3,4}$, 
Luke Shingles$^{5,39}$, 
Ken W. Smith$^{5}$,
James Steiner$^{40}$,\\
Richard Wainscoat$^{20}$,
Thomas Wevers$^{16}$,
Sheng Yang$^{22}$\\
{\small $^{\bf 1}$Kavli Institute for Astrophysics and Space Research,}\\ {\small Massachusetts Institute of Technology, Cambridge, MA, USA}\\
{\small $^{\bf 2}$Department of Astrophysics/IMAPP, Radboud University,}\\
{\small PO Box 9010, 6500 GL, The Netherlands}\\
{\small $^{\bf 3}$Institute of Gravitational Wave Astronomy, University of Birmingham, B15 2TT, UK}\\
{\small $^{\bf 4}$School of Physics and Astronomy, University of Birmingham, B15 2TT, UK}\\
{\small $^{\bf 5}$Astrophysics Research Centre, School of Mathematics and Physics,}\\ 
{\small Queen's University Belfast, Belfast, BT7 1NN, UK}\\
{\small $^{\bf 6}$International Centre for Radio Astronomy Research,} \\
{\small Curtin University, GPO Box U1987, Perth, WA 6845, Australia}\\
{\small $^{\bf 7}$Center for Interdisciplinary Exploration and Research in Astrophysics (CIERA)
and Department of}\\{\small Physics and Astronomy, Northwestern University, 1800 Sherman Ave, Evanston, IL 60201, USA}\\
{\small $^{\bf 8}$Astrophysics, Department of Physics, University of Oxford, Keble Road, Oxford OX1 3RH, UK}\\
{\small $^{\bf 9}$Inter-University Centre for Astronomy and Astrophysics, Pune, India}\\
{\small $^{\bf 10}$NASA Goddard Space Flight Center, Greenbelt, MD, USA}\\ \and
{\small $^{\bf 11}$Department of Physics, Syracuse University, Syracuse, New York, USA}\\
{\small $^{\bf 12}$Racah Institute of Physics, The Hebrew University of Jerusalem, Jerusalem 91904, Israel}\\
{\small $^{\bf 13}$Leiden Observatory, Leiden University, Postbus 9513, 2300 RA, Leiden, The Netherlands }\\
{\small $^{\bf 14}$Department of Physics and Astronomy, Johns Hopkins University,}\\{\small 3400 N. Charles St., Baltimore MD 21218, USA}\\
{\small $^{\bf 15}$Department of Physics \& Astronomy. University of Southampton, }\\{\small Southampton SO17 1BJ, UK}\\
{\small $^{\bf 16}$European Southern Observatory, Alonso de C\'ordova 3107, Casilla 19, Santiago, Chile}\\
{\small $^{\bf 17}$The School of Physics and Astronomy, Tel Aviv University, Tel Aviv 69978, Israel}\\
{\small $^{\bf 18}$CIFAR Azrieli Global Scholars program, CIFAR, Toronto, Canada}\\
{\small $^{\bf 19}$School of Physics, O’Brien Centre for Science North, University College Dublin,}\\ {\small Belfield, Dublin 4, Ireland}\\
{\small $^{\bf 20}$Institute for Astronomy, University of Hawaii}\\
{\small $^{\bf 21}$DTU Space, National Space Institute, Technical University of Denmark,}\\ {\small Elektrovej 327, 2800 Kgs. Lyngby, Denmark}\\
{\small $^{\bf 22}$The Oskar Klein Centre, Department of Astronomy, Stockholm University,}\\{\small AlbaNova, SE-10691 Stockholm, Sweden}\\
{\small $^{\bf 23}$Instituto de Astrof\'{\i}sica, Pontificia Universidad Cat\'olica,}\\{\small Vicu\~na Mackenna 4860, 7820436 Santiago, Chile}\\
{\small $^{\bf 24}$Millennium Institute of Astrophysics, Nuncio Monse\~nor~S\'otero Sanz
100,}\\{\small Of. 104, Providencia, 7500000 Santiago, Chile}\\
{\small $^{\bf 25}$IAP/Paris \& Sorbonne University}\\
{\small $^{\bf 26}$Institute of Space Sciences (ICE, CSIC), Campus UAB, Carrer de Can Magrans,}\\{\small s/n, E-08193 Barcelona, Spain.}\\
{\small $^{\bf 27}$Institut d’Estudis Espacials de Catalunya (IEEC), E-08034 Barcelona, Spain}\\
{\small $^{\bf 28}$Astronomical Observatory, University of Warsaw, Al. Ujazdowskie 4,}\\{\small 00-478 Warszawa, Poland}\\
{\small $^{\bf 29}$Research Computing, ITS Division, Northeastern University}\\
{\small $^{\bf 30}$Department of Physics, University of Warwick, Gibbet Hill Road, Coventry CV4 7AL, UK}\\
{\small $^{\bf 31}$NASA Marshall Space Flight Center}\\
{\small $^{\bf 32}$Department of Astronomy, University of California, 501 Campbell Hall,}\\{\small Berkeley, CA 94720, USA}\\
{\small $^{\bf 33}$Tuorla Observatory, Department of Physics and Astronomy, University of Turku,}\\{\small FI-20014 Turku, Finland}\\
\and
{\small $^{\bf 34}$Graduate Institute of Astronomy, National Central University,}\\{\small 300 Jhongda Road, 32001 Jhongli, Taiwan
}\\
{\small $^{\bf 35}$INAF-Osservatorio Astronomico d'Abruzzo,
via M. Maggini snc,}\\{\small I-64100 Teramo, Italy}\\
{\small $^{\bf 36}$Instituto de Astrof\'isica de Andaluc\'ia (IAA-CSIC), }\\{\small Glorieta de la Astronom\'ia s/n, E-18008 Granada, Spain}\\
{\small $^{\bf 37}$Facultad de Ciencias, Universidad de Zaragoza, Pedro Cerbuna 12, E-50009 Zaragoza, Spain}\\
{\small $^{\bf 38}$School of Sciences, European University Cyprus,}\\
{\small Diogenes Street, Engomi, 1516 Nicosia, Cyprus}\\
{\small $^{\bf 39}$GSI Helmholtzzentrum f\"ur Schwerionenforschung, Planckstraße 1,}\\{\small 64291 Darmstadt, Germany}\\
{\small $^{\bf 40}$Smithsonian Astrophysical Observatory; 60 Garden Street Cambridge, MA 02138, USA}\\
{\small $^{\bf 41}$SRON, Netherlands Institute for Space Research, Niels Bohrweg 4, 2333 CA Leiden, The Netherlands}\\
{\small $^{\bf 42}$The Oskar Klein Centre, Department of Physics, Stockholm University,}\\{\small AlbaNova, SE-10691 Stockholm, Sweden}\\
}
\date{}
\begin{document} 
\baselineskip24pt
\maketitle 
\begin{sciabstract}
A black hole can launch a powerful relativistic jet after it tidally disrupts a star. If this jet fortuitously aligns with our line of sight, the overall brightness is Doppler boosted by several orders of magnitude. Consequently, such on-axis relativistic tidal disruption events (TDEs) have the potential to unveil cosmological (redshift $z>$1) quiescent black holes and are ideal test beds to understand the radiative mechanisms operating in super-Eddington jets. Here, we present multi-wavelength (X-ray, UV, optical, and radio) observations of the optically discovered transient \target at $z=1.193$. Its unusual X-ray properties, including a peak observed luminosity of $\gtrsim$10$^{48}$ erg\,s$^{-1}$, systematic variability on timescales as short as 1000 seconds, and overall duration lasting more than 30 days in the rest-frame are traits associated with relativistic TDEs. The X-ray to radio spectral energy distributions spanning 5-50 days after discovery can be explained as synchrotron emission from a relativistic jet (radio), synchrotron self-Compton (X-rays), and thermal emission similar to that seen in low-redshift TDEs (UV/optical). Our modeling implies a beamed, highly relativistic jet akin to blazars but requires extreme matter-domination, i.e, high ratio of electron-to-magnetic field energy densities in the jet, and challenges our theoretical understanding of jets. 
%We disfavour a scenario in which X-rays are produced by inverse Compton scattering of external thermal photons from the inferred optical/UV blackbody as it requires fine tuning.

% Our work demonstrates the importance of prompt multiwavelength monitoring observations of TDEs to disentangle the many physical processes that can occur within newly launched highly relativistic jets. 

\end{sciabstract}

\target was discovered in the optical waveband by the Zwicky Transient Facility (ZTF; \cite{ztf}) on 11 February 2022 as a fast-evolving transient, and was publicly reported to the Gamma-ray Coordination Network (GCN) on 14 February 2022 \cite{cmcdiscovery:Igor}. We confirmed the rapid evolution of this transient in the Asteroid Terrestrial-impact Last Alert System (ATLAS) survey data with a non-detection 24~hrs before the ZTF discovery and a subsequent decline of 0.6 magnitudes per day \cite{atlasLC:Fulton}.
A radio counterpart was identified in Karl G. Jansky Very Large Array (VLA) observations on 15 February 2022 \cite{firstradiogcn}. While the optical spectrum taken on 16 February 2022 revealed a featureless continuum \cite{gmosspectrum}, spectral features were detected in subsequent spectra taken one day later with the European Southern Observatory's (ESO) Very Large Telescope (VLT; \cite{cmcredshift}) and Keck/DEIMOS \cite{cmcredshift2}. In particular, the detection of [OIII] $\lambda$5007 emission and CaII, MgII and FeII absorption lines yielded a redshift measurement of $z=1.193$ or luminosity distance of 8.45 Gpcs \cite{cmcredshift,cmcredshift2}. The source did not have a neutrino counterpart\cite{2022ATel15239....1P}. Our follow-up X-ray (0.3--5~keV) observations with the Neutron star Interior Composition ExploreR (\nicer) on 16 February 2022 revealed a luminous X-ray counterpart \cite{cmcfirstnicer}. We also triggered additional multi-wavelength observations with numerous facilities, including {\it AstroSat} and The Neil Gehrels \swift\ Observatory (\swift) in the X-rays and the UV (see Extended Data Figures \ref{fig:xrtimage} and  \ref{fig:xrayevol}). We obtained an optical spectrum with ESO/VLT (Extended Data Figure \ref{fig:opticalspectrumvlt}) and imaging with several optical telescopes. In the radio band, we acquired multi-frequency data with the VLA, the Arcminute Microkelvin Imager-Large Array (AMI-LA) and the European Very Long Baseline  Interferometry (VLBI) Network (EVN; see ``Observations and Data Analysis'' in Methods for details on these observations). We adopt Modified Julian Date (MJD) 59621.4458 (the discovery epoch) as the reference time throughout the paper and all relative times are in the observer frame unless otherwise mentioned.

\begin{figure}[ht]
\begin{center}
%\hspace{-0.35in}
\includegraphics[width=0.95\textwidth, angle=0]{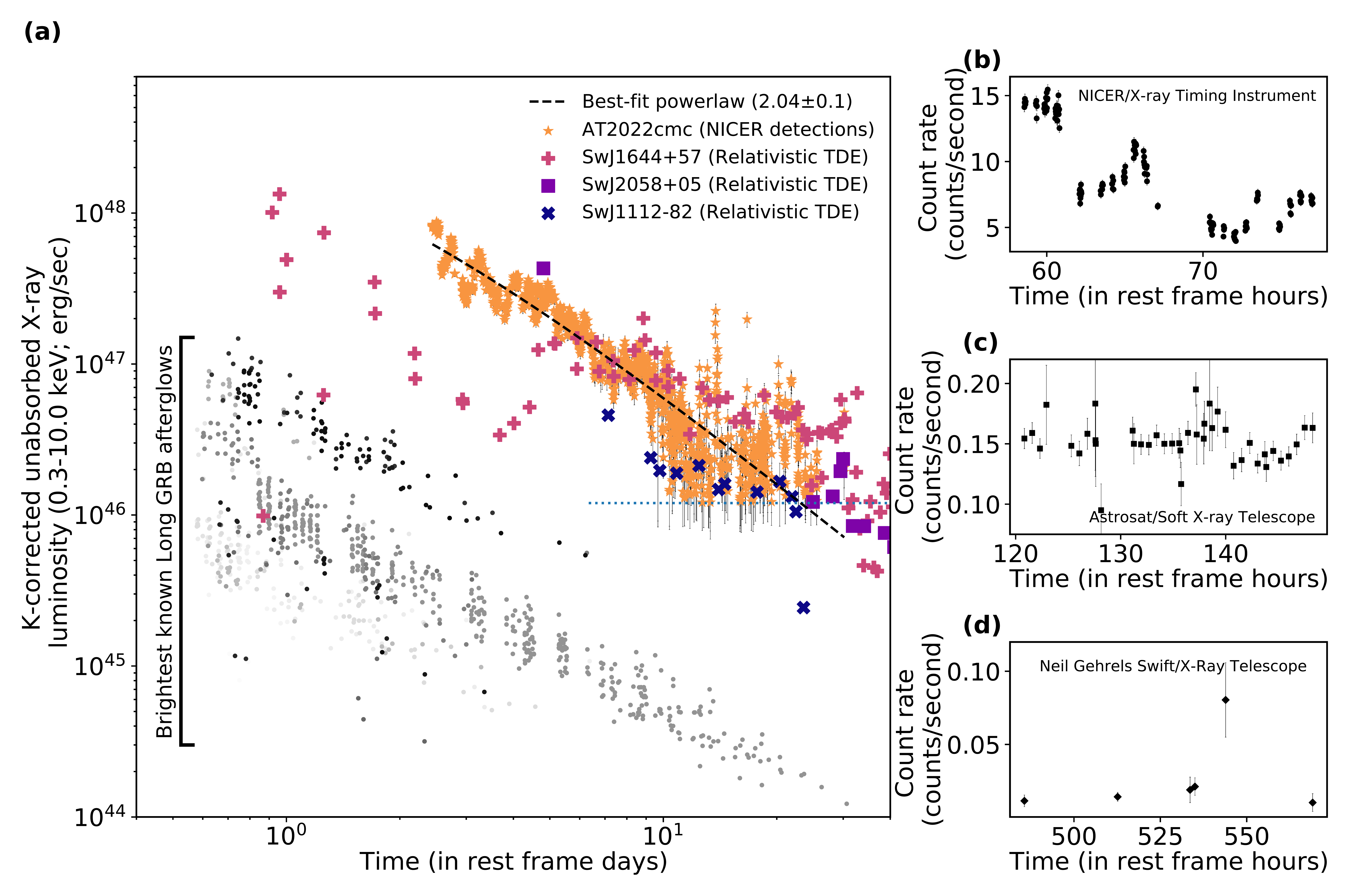}
\end{center}
%\vspace{-.35cm} 
\caption{{\bf \target's X-ray evolution on various timescales at different epochs.} {\bf (a) \target's k-corrected unabsorbed 0.3-10 keV X-ray luminosity (filled orange stars) in comparison to the most luminous known X-ray transients.} The filled circles with different shades of grey are a sample of 56 of the most luminous GRB X-ray afterglows known \cite{Gompertz18}. Only data past 50,000 rest-frame seconds is shown to highlight the late time emission from these afterglows. \target is significantly more luminous than any known GRB afterglow and its X-ray luminosity is only comparable to  previously-known relativistic jetted TDEs Sw~J1644+57 (filled green crosses), Sw~J2058+05 (filled cyan squares) and Sw~J1112-82 (filled purple Xs). The dotted horizontal blue line at 1.2$\times$10$^{46}$ erg s$^{-1}$ is an estimate of \nicer's background-limited sensitivity limit for sources at $z=1.193$. See ``GRB and TDE Comparison Data'' in Methods for a description of the comparison sample used in this Figure. {\bf (b) \target's sample \nicer~ (0.3-5 keV) light curve} highlighting variability on hours timescale (also see Extended Data Figure~\ref{fig:figpds}). {\bf (c) \target's Astrosat (0.5-7 keV) light curve showing variability on hours timescale.} {\bf (d) \target's \swift\ X-ray (0.3-8 keV) light curve} highlighting a flare more than 3 weeks (in rest-frame) after initial discovery. All the light curves are background-corrected. In panels (b)-(d), background-corrected count rates (counts~s$^{-1}$) vs time in rest frame hours since MJD 59621.4458 are shown. All the errorbars represent 1$\sigma$ uncertainties. These data are provided as supplementary files. }\label{fig:fig1}
\end{figure}

\begin{figure}[ht]
\begin{center}
%\hspace{-1.35cm}
\includegraphics[width=\textwidth, angle=0]{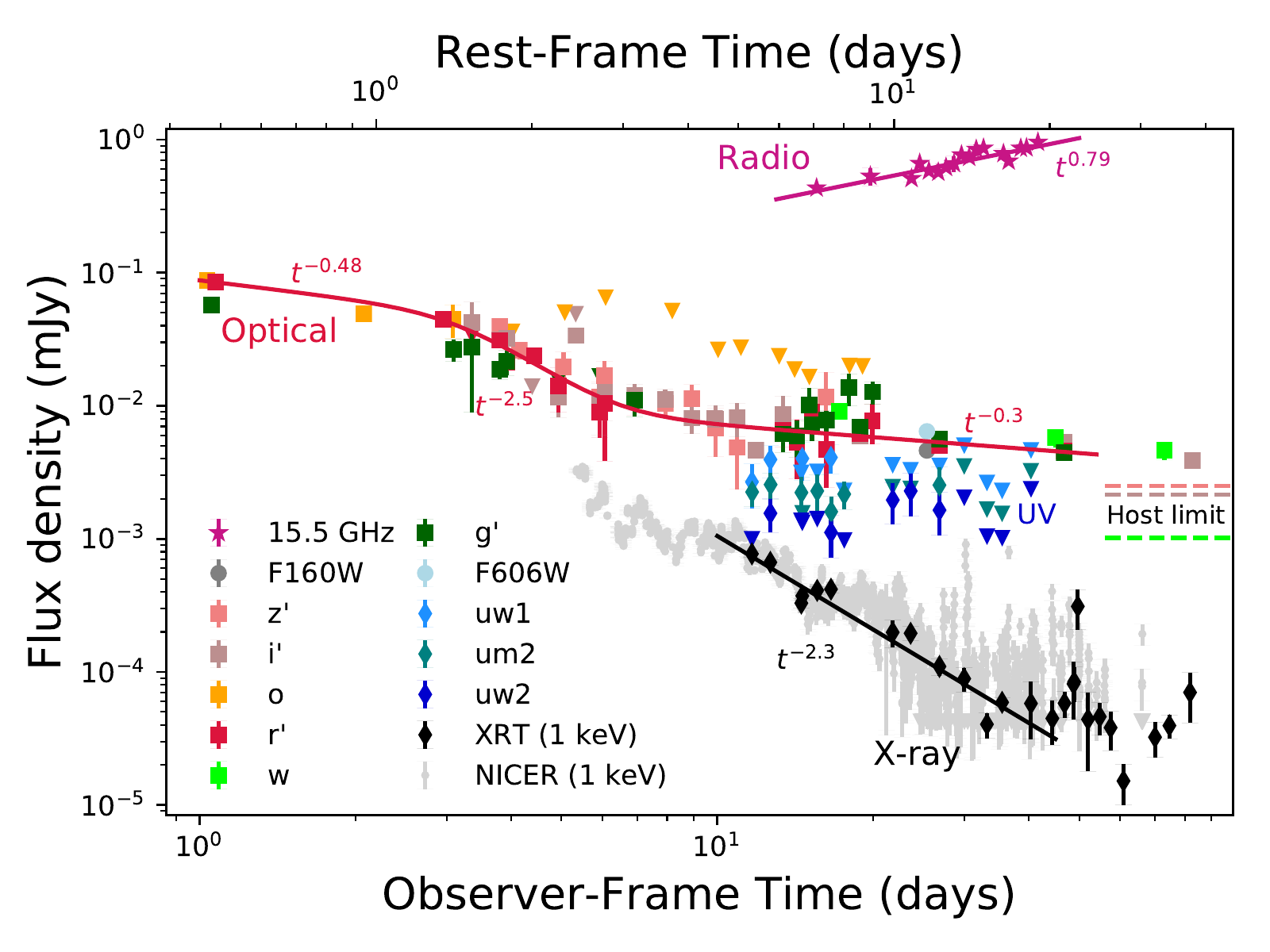}
\end{center}
%\vspace{-.35cm} 
\caption{{\bf \nicer\ (small grey points), \swift/XRT and UVOT (diamonds), \hst\ (circles), ground-based optical (squares), and radio (stars) light curves of \target\ spanning from $\approx1$--83~days after discovery}, together with single / smoothly broken power-law models fit to the \swift/XRT (black), $r^\prime$-band (red) and 15~GHz (violet) light curves with the corresponding best-fit indices indicated. The \swift\ and \nicer\ X-ray light curves have been converted from 0.3--5\,keV observer frame observed flux to flux density at 1~keV using the average and time-resolved X-ray spectral fits, respectively  (Section~\ref{data:xrt} and \ref{supsec:nicer}). The optical light curve exhibits a steep decay at $\approx1$--3 days in the rest frame, followed by a plateau, during which the radio light curve is seen to rise. Dashed lines indicate $w$, $i$, and $z$-band upper limits on underlying host emission obtained from deep stacks of PanSTARRS pre-discovery images (see ``Constraints on host luminosity'' and Extended Data Figure \ref{fig:legacyimage} in Methods). Upper limits are indicated by inverted triangles. All the photometry presented in this figure represents observed values that are corrected for Galactic extinction. This data is available as a supplementary file (Extended Data Table \ref{tab: allphot}). The multi-frequency VLA SED taken on 2022 February 27 is shown instead in Fig.~\ref{fig:SEDs}. All the errorbars represent 1$\sigma$ uncertainties.}\label{fig:multiband_lcs}
\end{figure}

\target's most striking property is its high isotropic peak X-ray luminosity of $\gtrsim10^{48}$ erg\,s$^{-1}$ (orange data points in panel (a) of Figure \ref{fig:fig1}). High apparent luminosity can be caused by gravitational lensing, however this contributes no more than a 10\% enhancement for \target (see ``Estimate of gravitational lens magnification by a foreground structure" in Methods). \target's second compelling aspect is its rapid X-ray variability over a wide range of timescales: during the weeks after initial optical discovery, it showed variability on timescales ranging from 1000~s to many days (see panels (a)--(d) of Figure \ref{fig:fig1}, Extended Data Figure \ref{fig:figpds}, and ``Shortest X-ray variability timescale'' in Methods). The X-ray spectrum is generally consistent with a simple power law model with the best-fit photon index varying between 1.3-1.9 (Extended Data Figure \ref{fig:xrayevol} and Extended Data Table \ref{tab: xraydata}). There are intermittent rapid flares during which the X-ray spectrum deviates from a power law model (see ``$\gamma$-rays and X-rays/\nicer'' in Methods). \target's observed optical and UV light curves exhibit three phases after reaching their peaks: an early slow decline\footnote{We use the convention, $F_\nu(\nu)\propto t^\alpha\nu^\beta$ throughout, where $F_\nu$ is the flux per unit frequency, $\nu$ is the observed frequency, $\alpha$ is the temporal decay rate, and $\beta$ is the spectral index.} phase at $\lesssim3.1$~days with a decline rate $\alpha\approx-0.5$ steepening further to $\alpha\approx-2.5$ at $\approx6.4$~days, followed by a shallow decline ($\alpha\approx-0.3$) at $\gtrsim6.4$~days (see Figure \ref{fig:multiband_lcs}). An optical spectrum taken at $\approx15$~days shows a featureless blue continuum, which can be fit using a thermal model with a rest-frame temperature $\approx$3$\times$10$^{4}$ K (see Extended Data Figure \ref{fig:opticalspectrumvlt}). The 15~GHz flux density, on the other hand, has been rising monotonically with time at $\gtrsim10$~days (see Figure \ref{fig:multiband_lcs}). The radio spectrum appears to be consistent with the standard synchrotron self-absorption process from a single-emitting region (e.g., see \cite{radiosynch}).

\begin{figure}[ht]
\begin{center}
%\hspace{-1.35cm}
%\includegraphics[width=\textwidth, angle=0]{figures/AT2022cmc_SEDs.pdf}
\includegraphics[width=0.75\textwidth, angle=0]{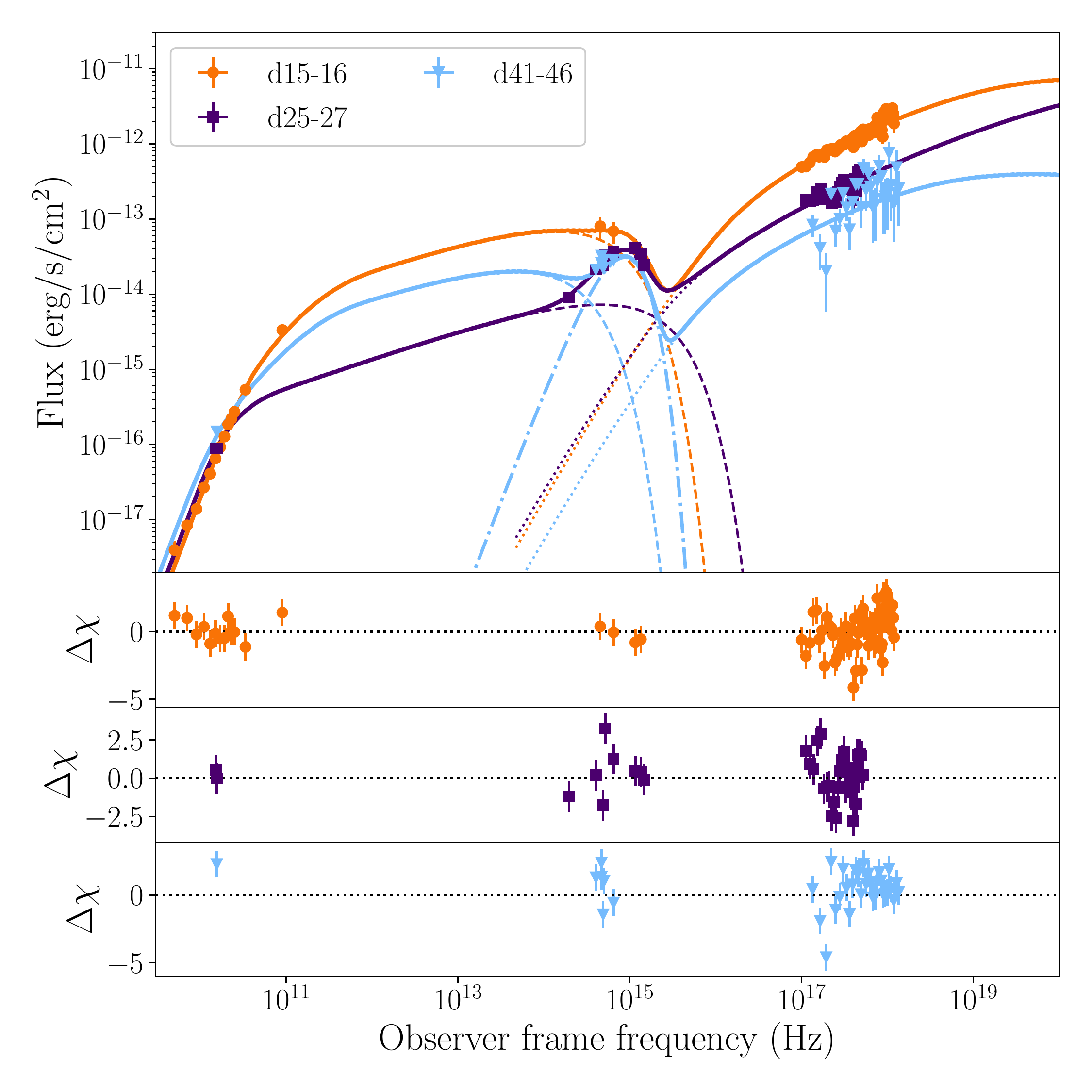}
\end{center}
%\vspace{-.35cm} 
\caption{{\bf \target's Multi-wavelength SEDs and their best-fit models}. SEDs from three epochs (times given as days post discovery) are fitted with a single-zone jet model comprising synchrotron (dashed), synchrotron self-Compton (dotted), and black body (dash-dot) emission components. The radio data are consistent with optically-thick synchrotron emission, while the X-ray emission is well fit by SSC originating from the same emitting region. The strength of the SSC component implies a strongly matter-dominated jet, with $U_e/U_B \geq 10^{2}$. The optical data at 25-27 and 41-46 days after discovery exhibit an excess over the synchrotron+SSC model; as a result, we added a black body component of temperature $T_{\rm bb} = 2.3\times10^{4}\,\rm{K}$ (measured in the source frame) and luminosity $L_{\rm bb}=1.7\times10^{45}\,\rm{erg/s}$. The corresponding radius is $R_{\rm bb} = 2.8\times10^{15}\,\rm{cm}$. Because of lack of optical/UV constraints on day 15-16, this component is assumed to remain constant between day 15-46 (see ``Multi-wavelength SED modeling'' and Extended Data Table \ref{tab: SED_fits} in Methods for more details). The data in this figure are available as a supplementary file. All the errorbars represent 1$\sigma$ uncertainties.}
\label{fig:SEDs}
\end{figure}

\target's high apparent X-ray energy output, extreme luminosity variations (a factor of $\sim$500 over a few weeks; see Figure \ref{fig:multiband_lcs} gray and black points) and fast variability requires an active central engine. Such an engine can be naturally explained by an extreme accretion episode onto a black hole which could be due to a stellar tidal disruption \cite{Rees1988}. Indeed, among transients, \target's apparent X-ray luminosity and evolution are only comparable to Sw~J1644+57 (e.g., \cite{Bloom2011}), Sw~J2058.4+0516 (e.g., \cite{SwJ2058:cenko+2011,Swj2058:pasham+2015}) and Sw~J1112.2-8238 \cite{SwJ1112:brown+2015}, the three TDEs with relativistic jets. 
%The full model fit yields a blackbody temperature of 2.3$\times$10$^{4}$ K (consistent with the  measurement of $\sim$3$\times$10$^{4}$ K from the VLT spectrum, after accounting for synchrotron contribution; see Extended Data Figure~\ref{fig:opticalspectrumvlt}). Similar thermal optical emission 
\target's thermal optical emission with temperature of $\sim$2.3$\times$10$^{4}$ K is often seen in low-redshift ($z\lesssim0.2$) TDEs \cite{ztfsample1} and could be from a newly formed accretion disk (e.g., \cite{Wevers2019_18fyk}), reprocessing (e.g., \cite{at2019qiz:2020MNRAS.499..482N}), or from debris stream self-collisions (e.g., \cite{2017ApJ...837L..30P, Piran2015}). The high optical/UV luminosity of $\approx2\times$10$^{45}$ erg~s$^{-1}$ at day 15-16 post-discovery (Figure \ref{fig:SEDs}) is only comparable to the extreme TDE candidate ASASSN-15lh \cite{Leloudas:2016NatAs...1E...2L}.
Based on the rich literature on accretion-driven outbursts from stellar-mass black holes in X-ray binaries, we now know that accretion and consequently related ejection can lead to variability on a wide range of timescales (see references in \cite{mcclin}). Thus, accretion/ejection following a tidal disruption could also naturally explain \target's observed flux variability over a wide range of timescales. 

Given the similar X-ray luminosity and variability to Sw~J1644+57, the best-studied TDE with a relativistic jet, we modelled \target's data under the jet paradigm. In a standard jet scenario, the radio through infrared/optical/UV data is dominated by non-thermal synchrotron emission \cite{Giannios2011,Romero17}. However, extrapolating \target's radio/optical/UV data to higher frequencies does not provide emission consistent with the observed X-ray flux (see ``Preliminary Considerations'' in Methods and Extended Data Figure~\ref{fig:SED_analytical}), suggesting that the high energy emission originates from a second component. Similar to blazars, this second component could naturally arise from inverse Compton scattering of either local synchrotron photons (synchrotron self-Compton, or SSC for brevity), or photons originating outside of the jet (external Compton, or EC). In both cases, the photons would interact with the electrons in the jet. Therefore, we investigated these scenarios by fitting three observed time-averaged spectral energy distributions (SEDs) with good multi-wavelength coverage (days 15-16, 25-27, and 41-46) with a simple jet model, consisting of a spherical, homogeneous, emitting region, similar to the approach commonly used to infer the properties of the emitting region in blazars \cite{Ghisellini09,Boettcher13,Tavecchio16}. The rapid X-ray variability on tens of minutes timescale and self-absorbed radio spectrum indicate that the observed radio and X-ray emission originate from a compact region rather than in an extended outflow, further motivating our single-zone approximation. 

We tested two emission models, one in which the only radiative mechanisms considered are synchrotron and SSC (model 1), and one including EC of thermal photons originating outside of the jet (model 2). Model 1 (the synchrotron+SSC model), shown in Figure \ref{fig:SEDs}, provides an acceptable fit to the radio through the X-ray SEDs ($\chi^{2}/\rm{d.o.f.}=2.2$), albeit with extreme parameters (see below); model 2 on the other hand is disfavored because it cannot explain the radio flux, while still resulting in similarly extreme parameters (see ``Modeling results'' in Methods). The best-fitting parameters for both models are reported in Extended Data Table \ref{tab: SED_fits}. We caution that these numbers could change significantly with a more complex and physical model, and the fits presented here purely constitute a check that the data is consistent with the emission from a relativistic jet. 

The main trend emerging from model 1 is that the jet has to be very powerful ($\approx 10^{46-47}$ $\rm{erg\,s}^{-1}$, depending on its composition) and strongly beamed: the Doppler factor is $\delta=[\Gamma_{\rm j}(1-\beta_{\rm j}\cos(\theta)]^{-1}\approx 100$, where $\Gamma_{\rm j}\approx 86$ is the jet bulk Lorentz factor, $\beta_{\rm j}$ the corresponding speed in units of the speed of light, and $\theta$ is the jet viewing angle. On the other hand, model 2 requires somewhat lower jet power ($\approx 10^{45}$ $\rm{erg\,s}^{-1}$), and a smaller bulk Lorentz factor $\Gamma_{\rm j}\approx 5$ and Doppler factor $\delta\approx10$. Under the jet paradigm, the observed X-rays and their variability arise from within the jet; as a result, a size constraint can be compared to the observed variability timescale in order to check for consistency. Based on a simple causality argument, we require the size of the emitting region to be smaller than the minimum variability timescale$\times$speed of light$\times$Doppler factor $\approx1000$~s $\times$ 3$\times$10$^{10}\times\delta$ cm $\approx3\times10^{13}\times\delta$ cm for our case, where the factor $\delta$ accounts for relativistic beaming \cite{RadiativeBook}. The emitting region inferred has an estimated radius of $\approx 10^{15-16}\,\rm{cm}$ from model 1 and $\approx10^{14}\,\rm{cm}$ from model 2. Both of these estimates are consistent with the hour-long variability timescale observed by \nicer but are only marginally consistent with $\sim$1000~s X-ray variations. Such rapid variability has also been observed in some extreme blazar flares (e.g.,  \cite{Aharonian07,Hayashida15}), and is inconsistent with the simple homogeneous, time-independent single-zone model presented here. Instead, it can be reproduced using a complex in-homogeneous, time-dependent model \cite{Raiteri17}. However, applying such a model to \target  is beyond the scope of this work. 

\begin{figure}[ht]
\begin{center}
\includegraphics[width=\textwidth, angle=0]{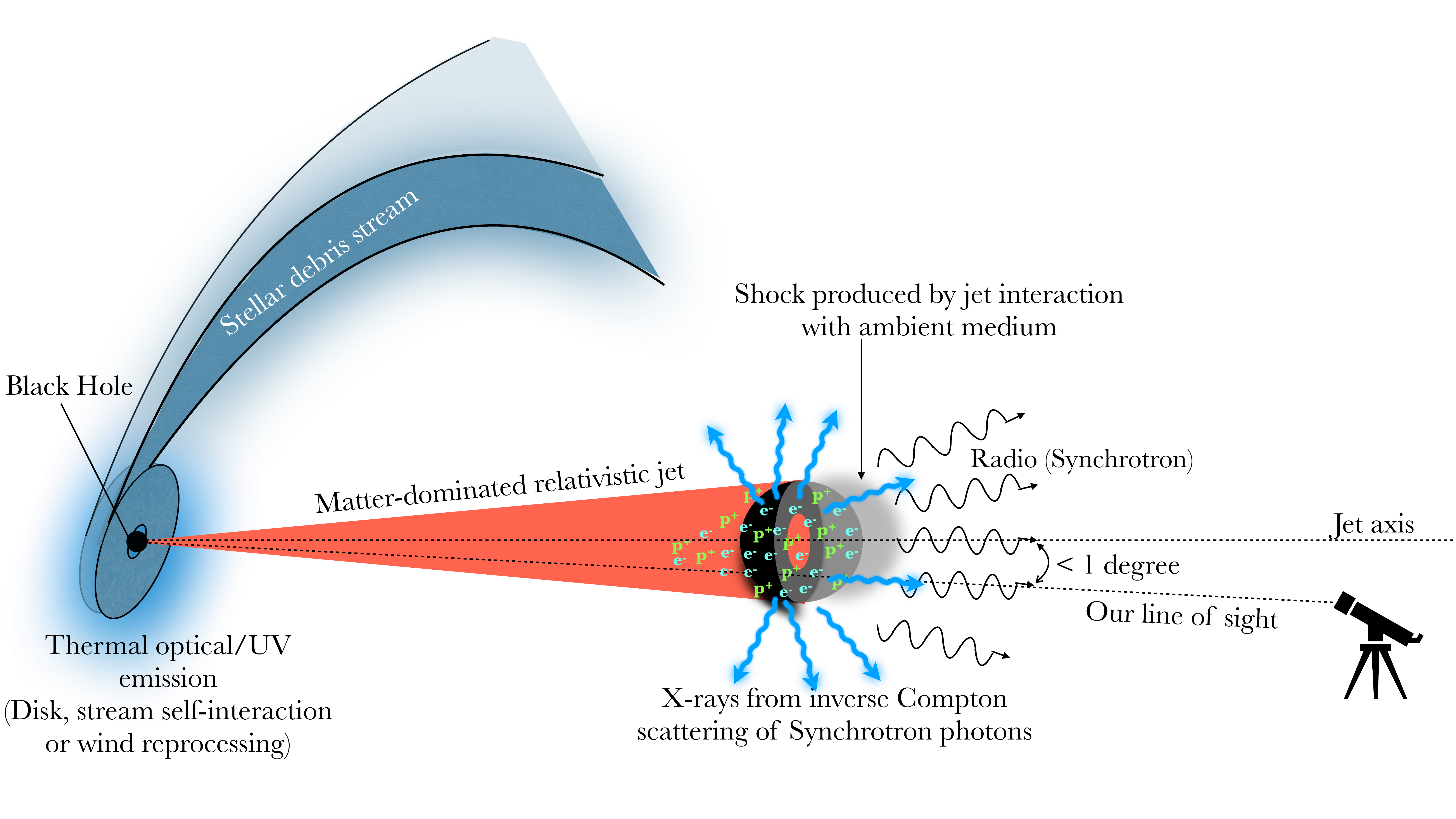}
\end{center}

\caption{{\bf Schematic of our proposed scenario for \target}. A mass-loaded, highly relativistic jet with a bulk Lorentz factor $\sim$80 can explain \target's  multi-wavelength SED with radio emission originating from synchrotron processes and X-rays from SSC (see ``Multi-wavelength SED modeling'' and Extended Data Table \ref{tab: SED_fits} in Methods). The optical/UV emission part of the SED on day 25 is consistent with thermal emission with a temperature of $\sim$2.3$\times$10$^{4}$ K and luminosity of 2$\times$10$^{45}$ erg~s$^{-1}$ (rest-frame). These are comparable to low-z non-jetted TDEs \cite{ztftdesample2}. It could originate from an accretion disk, reprocessing by an outflow (e.g., \cite{at2019qiz:2020MNRAS.499..482N}) or from stellar debris stream self-collisions \cite{Piran2015}. Our viewing angle with respect to the jet-axis is estimated from our SED modeling to be $<1$ degrees (see Extended Data Table \ref{tab: SED_fits}). }
\label{fig:schematic}
\end{figure}

Both models 1 and 2 require a strong SSC contribution to match the X-ray flux. In order for this to happen, we require a strongly matter-dominated jet, i.e., most of the power is carried by the electrons and protons within the jet, rather than by the magnetic field. Such a matter dominated flow is in tension with the common theoretical paradigm that jets are magnetically-dominated at their launching point, and then accelerate by turning the magnetic field into bulk kinetic energy until they reach rough equipartition \cite{McKinney06,Chatterjee19}, but is in line with \cite{Coughlin2020} who proposed a structured, radiation-driven jet powered by super-Eddington accretion. The jet collimation could be provided by the pressure of the surrounding accretion flow, which is highly inflated during the super-Eddington phase (e.g., \cite{Bromberg07, Kohler12, Coughlin14, Coughlin2020}). These issues are also often encountered when modelling blazar jets with a dominant SSC component, \cite{Tavecchio16,Costamante18}, as well as M87 \cite{M87mwlEht}, and likely points at the need for more complex models. A schematic of our proposed, albeit simple, model (synchrotron+SSC+thermal optical/UV) is shown in Figure \ref{fig:schematic}.

Finally, our SED  models imply that the underlying physics in \target's jet maybe  distinct compared to Sw~J1644+57 and Sw~J2058+05, as in those sources SSC cannot produce the observed X-ray emission \cite{crumley}. In Sw~J1644+57 it has been argued that the X-rays originate from a corona/base of a jet through external inverse Compton scattering by a photon field coming from either the disk (e.g., \cite{Bloom2011, seifina17}) or from the disk wind (e.g., \cite{crumley}). This external inverse Compton model has also been successfully applied to Sw~J2058+05 \cite{swj2058EIC, seifina17}. Instead, in \target EC cannot explain the observed X-rays (see ``Modeling results'' in Methods), and thus its high energy emission appears to be driven by different mechanisms compared to previous relativistic TDEs.

While our models provide strong evidence that the multi-wavelength emission of \target is powered by a relativistic jet, they also show that a more complex model is required to probe the physics of the jet self-consistently. The data presented in this paper provide an unprecedented opportunity to explore detailed jet physics at extreme mass accretion rates.

As a relativistic jet is able to explain the multi-wavelength properties of \target, we now investigate the plausible mass of the black hole engine. At the low mass end, $\sim$10 $M_{\odot}$, the most powerful known jets are launched following Gamma Ray Bursts (GRBs). A GRB afterglow interpretation can be ruled out due to the: 1) unusually high  X-ray luminosity, 2) fast variability out to weeks after discovery, 3) overall duration of \target, and 4) non-synchrotron SED (see ``Arguments against a GRB afterglow'' in Methods for a more thorough/detailed discussion). We disfavour a blazar flare/outburst for three reasons. First, the light curves of blazar flares show stochastic variability on top of a fairly constant, low flux (e.g. \cite{Raiteri17}), while \target shows a smooth decay structure typical of transients powered by a sudden (and possibly subsequently sustained) deposition of energy. Second, all blazar classes have a flat radio spectrum, $F(\nu)\propto \nu^{0}$, while \target exhibits a strongly self-absorbed spectrum with $F(\nu)\propto \nu^{2}$. Finally, a large amplitude optical brightness enhancement of $\sim$4 magnitudes (see ``Constraints on host luminosity'' in Methods and supplementary data) is unusual for blazars (e.g., compare with  \cite{Raiteri17}). In addition to this, there is no gamma-ray source detected by Fermi/LAT within $1^{\rm o}$ diameter from \target. 

A TDE is largely characterized by the pericenter distance (the closest approach between the star and the black hole), the stellar properties, and the black hole mass. The pericenter distance does not affect the accretion rate if the disruption is full (e.g., \cite{Lacy1982, Guillochon2013, Stone2013, Norman2021}), while if it is partial there is a steep falloff in luminosity with increasing distance (e.g., \cite{Guillochon2013, TDEpartialpowlaw:Coughlin+2019, Nixon2021}). For a star of radius $R_{\star}$ and mass $M_{\star}$ %, and dynamical time $T_{\star} = R_{\star}^{3/2}/\sqrt{GM_{\star}}$ 
and a black hole of mass $M$, the characteristic %fallback timescale for a TDE is $T = T_{\star}(M/M_{\star})^{1/2}$ \cite{Rees1988}, and hence the 
TDE accretion rate is $\propto (M_{\star}/R_{\star})^{3/2}(M/M_{\star})^{-1/2}$. For a main sequence star with $R_{\star} \propto M_{\star}$ the luminosity is therefore $\propto  M_{\star}^{1/2}$, and a very massive (and rare) star is needed to substantially modify the accretion rate (e.g., Figure 4 of \cite{Golightly2019}). On the other hand, the Eddington ratio for a TDE scales as $M^{-3/2}$, and a modest decrease in black hole mass yields a large increase in the Eddington fraction. Given these considerations and the approximate scaling of the X-ray luminosity as $\propto t^{-9/4}$ \cite{TDEpartialpowlaw:Coughlin+2019}, we suggest that \target could have been powered by the partial disruption (near the full disruption threshold) of a dwarf star by a relatively low-mass black hole and its super-Eddington accretion.

% However, both an intermediate-mass black hole weighing a few$\times$(10$^{2-5}$) M$_{\odot}$ disrupting a white dwarf and a supermassive black hole (a few$\times$10$^{6-8}$) disrupting a solar type star remain as possible scenarios for powering \target. The observed high apparent X-ray luminosity implies that the emission is highly super-Eddington in both of these cases. {\bf The X-ray luminosity evolution when fit with a power-law model implies a slope between 2-2.3 depending on the epoch. This index value is consistent with 2.2 predicted from analytical theory for a partial TDE \cite{TDEpartialpowlaw:Coughlin+2019, microTDE:2016ApJ...823..113P}, and is comparable to the index inferred from X-ray data of SwJ2058+05  \cite{SwJ2058:cenko+2011}. It is thus possible that \target's emission may have been powered by a partial TDE. }

While non-relativistic TDEs are now routinely discovered (roughly one every few weeks) in the nearby Universe (redshift, $z\lesssim$ 0.2) \cite{ztfsample1, ztftdesample2}, Doppler-boosted TDEs such as \target\ can push the redshift barrier as they are orders of magnitude more luminous. \target's multi-wavelength properties are consistent with a TDE with a relativistic jet closely aligned with our line of sight. This makes \target the farthest TDE known to-date. It is also the  first relativistic TDE to be identified in over 11 years \cite{Brown2015}, and the first such event to be identified by an optical sky survey. All these factors bolster the exciting prospect of unveiling $z>1$ TDEs and consequently black holes in the upcoming  era of {\it LSST/Rubin} observatory \cite{lssttdes}.

% %%%%%%%%%%%%%%%%%%%%%%%%%%%%%%%%%%%%%%%%%%%%%%%%%%%%%%%%%%%%%%%%%%%%%%%%%%%%%%%%%%%%%%%%%%%%%%
% % ---- Figure--- Figure ---- Figure--- Figure ---- Figure--- Figure --- Figure--- Figure ----%
% %%%%%%%%%%%%%%%%%%%%%%%%%%%%%%%%%%%%%%%%%%%%%%%%%%%%%%%%%%%%%%%%%%%%%%%%%%%%%%%%%%%%%%%%%%%%%%

\clearpage
\vfill\eject

% %%%%%%%%%%%%%%%%%%%%%%%%%%%%%%%%%%%%%%%%%%%%%%%%%%%%%%%%%%%%%%%%%%%%%%%%%%%%%%%%%%%%%%%%%%
% %%%%%%%%%%%%%%%%% muliband light curves %%%%%%%%%%%%%%%%%%%%%%%%%%%
% %%%%%%%%%%%%%%%%%%%%%%%%%%%%%%%%%%%%%%%%%%%%%%%%%%%%%%%%%%%%%%%%%%%%%%%%%%%%%%%%%%%%%%%%%%

\newpage
\vfill\eject

% %%%%%%%%%%%%%%%%%%%%%%%%%%%%%%%%%%%%%%%%%%%%%%%%%%%%%%%%%%%%%%%%%%%%%%%%%%%%%%%%%%%%%%%%%%
% %%%%%%%%%%%%%%%%% SEDs %%%%%%%%%%%%%%%%%%%%%%%%%%%
% %%%%%%%%%%%%%%%%%%%%%%%%%%%%%%%%%%%%%%%%%%%%%%%%%%%%%%%%%%%%%%%%%%%%%%%%%%%%%%%%%%%%%%%%%%

\newpage
\vfill\eject

%%%%%%%%%%%%%%%%%%%%%%%%%%%%%%%%%%%%%%%%%%%%%%%%%%%%%%%%%%%%%%%%%%%%%%%%%%%%%%%%%%%%%%%%%%
%%%%%%%%%%%%%%%%%%%%%%%%%%%%% Stacked XRT and late time MOS1 images %%%%%%%%%%%%%%%%%%%%%%
%%%%%%%%%%%%%%%%%%%%%%%%%%%%%%%%%%%%%%%%%%%%%%%%%%%%%%%%%%%%%%%%%%%%%%%%%%%%%%%%%%%%%%%%%%

\newpage
\vfill\eject

%%%%%%%%%%%%%%%%%%%%%%%%%%%%%%%%%%%%%%%%%%%%%%%%%%%%%%%%%%%%%%%%%%%%%%%%%%%%%%%%%%%%%%%%%%%%%%
% ---- References --- References ---- References--- References ---- References--- References %
%%%%%%%%%%%%%%%%%%%%%%%%%%%%%%%%%%%%%%%%%%%%%%%%%%%%%%%%%%%%%%%%%%%%%%%%%%%%%%%%%%%%%%%%%%%%%%

\bibliographystyle{Science}
\bibliography{references}

\clearpage

\singlespace
\begin{scilastnote}
\begin{sloppypar}
\item[] {\bf Acknowledgments.}\\ 

DRP would like to thanks Dr. Simon Dicker for sharing details of the GBT observations. DRP was supported by NASA grant 80NSSC22K0961 for this work.

SJB  would like to thank their support from Science Foundation Ireland and the Royal Society (RS-EA/3471).

S. Schulze acknowledges support from the G.R.E.A.T. research environment, funded by {\em Vetenskapsr\aa det},  the Swedish Research Council, project number 2016-06012.

FO acknowledges support from MIUR, PRIN 2017 (grant 20179ZF5KS) "The new frontier of the Multi-Messenger Astrophysics: follow-up of electromagnetic transient counterparts of gravitational wave sources" and the support of HORIZON2020: AHEAD2020 grant agreement n.871158.

GL and PC were supported by a research grant (19054) from VILLUM FONDEN. 

NCS acknowledge support by the Science and Technology Facilities Council (STFC), and from STFC grant ST/M001326/.

MN, BG, AA, ER and XS are supported by the European Research Council (ERC) under the European Union’s Horizon 2020 research and innovation programme (grant agreement No.~948381).

L. R. acknowledges the support given by the Science and Technology Facilities Council through an STFC studentship.

TL acknowledges support from the Radboud Excellence Initiative. 

TEMB acknowledges financial support from the Spanish 
Ministerio de Ciencia e Innovaci\'on (MCIN), the Agencia Estatal de 
Investigaci\'on (AEI) 10.13039\/501100011033 under the 
PID2020-115253GA-I00 HOSTFLOWS project, from Centro Superior de 
Investigaciones Cient\'ificas (CSIC) under the PIE project 20215AT016 
and the I-LINK 2021 LINKA20409, and the program Unidad de Excelencia 
Mar\'ia de Maeztu CEX2020-001058-M.

CCN thanks for funding from the Ministry of Science and Technology (Taiwan) under the contract 109-2112-M-008-014-MY3.

MPT acknowledges financial support from the State Agency for Research of the Spanish MCIU through the
"Center of Excellence Severo Ochoa" award to the Instituto de Astrofísica de Andalucía (SEV-2017-0709)
and through the grant PID2020-117404GB-C21 (MCI/AEI/FEDER, UE).

Support for AC was provided by ANID through grant ICN12\_12009 awarded to
the Millennium Institute of Astrophysics (MAS) and by ANID's Basal
projects AFB-170002 and FB210003.

ERC acknowledges support from the National Science Foundation through grant AST-2006684, and a Ralph E.~Powe Junior Faculty Enhancement Award through the Oakridge Associated Universities.

Pan-STARRS is a project of the Institute for Astronomy of the University of Hawaii, and is supported by the NASA SSO Near Earth Observation Program under grants 80NSSC18K0971, NNX14AM74G, NNX12AR65G, NNX13AQ47G, NNX08AR22G, 80NSSC21K1572  and by the State of Hawaii. 

This publication has made use of data collected at Lulin Observatory, partly supported by MoST grant 108-2112-M-008-001. We thank Lulin staff H.-Y. Hsiao, C.-S. Lin, W.-J. Hou and J.-K. Guo for observations and data management.

This work was supported by the Australian government through the Australian Research
Council's Discovery Projects funding scheme (DP200102471). 

The PanSTARRS1 Surveys (PS1) and the PS1 public science archive have been made possible through contributions by the Institute for Astronomy, the University of Hawaii, the Pan-STARRS Project Office, the Max-Planck Society and its participating institutes, the Max Planck Institute for Astronomy, Heidelberg and the Max Planck Institute for Extraterrestrial Physics, Garching, The Johns Hopkins University, Durham University, the University of Edinburgh, the Queen’s University Belfast, the Harvard-Smithsonian Center for Astrophysics, the Las Cumbres Observatory Global Telescope Network Incorporated, the National Central University of Taiwan, the Space Telescope Science Institute, the National Aeronautics and Space Administration under Grant No. NNX08AR22G issued through the Planetary Science Division of the NASA Science Mission Directorate, the National Science Foundation Grant No. AST–1238877, the University of Maryland, Eotvos Lorand University (ELTE), the Los Alamos National Laboratory, and the Gordon and Betty Moore Foundation.

RR and DP acknowledge partial support from the NASA Grant, 80NSSC19K1287, for contributions to NICER.

The European VLBI Network is a joint facility of independent European, African, Asian, and North American radio astronomy institutes. Scientific results from data presented in this publication are derived from EVN project code RM017A. e-VLBI research infrastructure in Europe is supported by the European Union’s Seventh Framework Programme (FP7/2007-2013) under grant agreement number RI-261525 NEXPReS.

A.H. is grateful for the support by the I-Core Program of the Planning and Budgeting Committee and the Israel Science Foundation, and support by ISF grant 647/18. This research was supported by Grant No. 2018154 from the United States-Israel Binational Science Foundation (BSF). We acknowledge the staff who operate and run the AMI-LA telescope at Lord's Bridge, Cambridge, for the AMI-LA radio data. AMI is supported by the Universities of Cambridge and Oxford, and by the European Research Council under grant ERC-2012-StG-307215 LODESTONE.

NICER is a 0.2-12 keV X-ray telescope operating on the International Space Station. The NICER mission and portions of the NICER science team activities are funded by NASA.

The {\it AstroSat} mission is operated by the Indian Space Research Organisation (ISRO), the data are archived at the Indian Space Science Data Centre (ISSDC). The  SXT data-processing software is provided by the Tata Institute of Fundamental Research (TIFR), Mumbai, India.  The UVIT data were checked and verified by the UVIT POC at IIA, Bangalore, India. 

MG is supported by the EU Horizon 2020 research and innovation programme under grant agreement No 101004719.

LJS acknowledges support by the European Research Council (ERC) under the European Union’s Horizon 2020 research and innovation program (ERC Advanced Grant KILONOVA No. 885281). 

MPT acknowledges financial support from the State Agency for Research of the Spanish MCIU through the
"Center of Excellence Severo Ochoa" award to the Instituto de Astrofísica de Andalucía (SEV-2017-0709)
and through the grant PID2020-117404GB-C21 (MCI/AEI/FEDER, UE).

Support for this work was provided by NASA through the Smithsonian Astrophysical Observatory (SAO) contract SV3-73016 to MIT for Support of the Chandra X-Ray Center (CXC) and Science Instruments.

S. Y. has been supported by the research project grant “Understanding the Dynamic Universe” funded by the Knut and Alice Wallenberg Foundation under Dnr KAW 2018.0067, and the G.R.E.A.T research environment, funded by {\em Vetenskapsr\aa det}, the Swedish Research Council, project number 2016-06012. 

SJS, SS, KWS, acknowledge funding from STFC Grant ST/T000198/1 and ST/S006109/1.

IA is a CIFAR Azrieli Global Scholar in the Gravity and the Extreme Universe Program and acknowledges support from that program, from the European Research Council (ERC) under the European Union’s Horizon 2020 research and innovation program (grant agreement number 852097), from the Israel Science Foundation (grant number 2752/19), from the United States - Israel Binational Science Foundation (BSF), and from the Israeli Council for Higher Education Alon Fellowship.

ECF is supported by NASA under award number 80GSFC21M0002.

The National Radio Astronomy Observatory is a facility of the National Science Foundation operated under cooperative agreement by Associated Universities, Inc.

GPS acknowledges support from The Royal Society, the Leverhulme Trust, and the Science and Technology Facilities Council (grant numbers ST/N021702/1 and ST/S006141/1).

L.G. acknowledges financial support from the Spanish Ministerio de Ciencia e Innovaci\'on (MCIN), the Agencia Estatal de Investigaci\'on (AEI) 10.13039/501100011033, and the European Social Fund (ESF) "Investing in your future" under the 2019 Ram\'on y Cajal program RYC2019-027683-I and the PID2020-115253GA-I00 HOSTFLOWS project, from Centro Superior de Investigaciones Cient\'ificas (CSIC) under the PIE project 20215AT016, and the program Unidad de Excelencia Mar\'ia de Maeztu CEX2020-001058-M.

{\bf Author contributions:} D.R.P led the overall project. 
{\bf Competing interests:} The authors
declare that there are no competing interests. {\bf Data and materials availability:} All the {\it NICER} and {\it Swift} data presented here is public and can be found in the NASA archives at the following URL: \url{https://heasarc.gsfc.nasa.gov/cgi-bin/W3Browse/w3browse.pl}. The multi-wavelength photometric values are provided as supplementary files. In addition, we also provide \nicer, {\it Astrosat/SXT}, \swift, and long-GRB  light curves used in Figure \ref{fig:fig1}.
\end{sloppypar}
\end{scilastnote}

\begin{scilastnote}
\item[] {\bf Supplementary Materials.}\\
Materials and Methods\\
Extended Data Figures 1 to 9\\
Extended Data Tables 1 to 3\\
% References (37-85)\\
Supplementary Text\\
\end{scilastnote}

%%%%%%%%%%%%%%%%%%%%%%%%%%%%%%%%%%%%%%%%%%%%%%%%%%%%%%%%%%%%%%%
%
%                   SUPPLEMENT SECTION
%
%%%%%%%%%%%%%%%%%%%%%%%%%%%%%%%%%%%%%%%%%%%%%%%%%%%%%%%%%%%%%%%%
\newpage
% \setcounter{page}{1}
% \renewcommand{\theequation}{S\arabic{equation}}
% \renewcommand{\thefigure}{\arabic{figure}}
% \renewcommand{\thetable}{\arabic{table}}
% \setcounter{figure}{0}

% \singlespace

%\noindent{\Huge {\bf Supplement}} \\[10 pt]

% \renewcommand{\figurename}{Extended Data Figure} 
\setcounter{page}{1}
\renewcommand{\theequation}{S\arabic{equation}}
\renewcommand{\figurename}{Extended Data Figure}
\renewcommand{\tablename}{Extended Data Table}
\setcounter{figure}{0}

% \noindent{\Huge {\bf Supplementary Information}} \\[10 pt]
\section*{{\Huge Methods.}}
%++++++++++++++++++++++++++++++++++++++++++++++++++++++++++++++++++++++++++++
% ------------------------------- DATA INTRODUCTION ------------------------
%++++++++++++++++++++++++++++++++++++++++++++++++++++++++++++++++++++++++++++

\section{\Large{\bf Observations and Data Analysis}}\label{supsec:data}
The data presented in this work was acquired by different telescopes/instruments across the electromagnetic spectrum. Below, we describe the data and the relevant reduction and analysis procedures. Throughout this paper, we
adopt a standard $\Lambda$CDM cosmology with H$_{0}$ = 67.4 km~s$^{-1}$~Mpc$^{-1}$, $\Omega_{m}$ = 0.315 and $\Omega_{\Lambda}$ = 1 - $\Omega_{m}$ = 0.685 \cite{planck}. Using the Cosmology calculator of \cite{Wright2006} \target's redshift of 1.193 corresponds to a luminosity distance of 8.45 Gpcs.

\subsection{\texorpdfstring{$\mathcal{}\gamma$}{Lg}-rays and X-rays}
\subsubsection{Fermi/LAT}
\target\ was not detected by {\it Fermi}/Large Area Telescope (LAT; 100 MeV to 10 GeV). During the 24 hour period starting on 27 February 2022 (UTC), i.e., days 15-16 after discovery, the upper limits on the photon flux and the energy flux are 2.76$\times$10$^{-7}$ photons cm$^{-2}$ s$^{-1}$, and 5.46$\times$10$^{-3}$ MeV cm$^{-2}$ s$^{-1}$, respectively.

\subsubsection{\textit{AstroSat}/SXT}
The {\it AstroSat} Soft X-ray Telescope (SXT; \cite{2017JApA...38...29S}) observed \target\ on 2022-02-23 for an exposure time of $52.8{\rm~ks}$ in the full window mode. We processed the level1 data using the SXT pipeline \textit{AS1SXTLevel2-1.4b} available at the Payload Operation Center (POC) website \footnote{\url{https://www.tifr.res.in/~astrosat\_sxt/sxtpipeline.html}}, and generated the orbit-wise cleaned event files which were then merged using the \textit{SXTMerger} tool\footnote{\url{https://github.com/gulabd/SXTMerger.jl}}. 
%We extracted the source spectrum and light curve using a circular region of radius $15^{\prime\prime}$ centered at the source position. We used the background spectrum and the redistribution matrix files available at the POC. We used an updated ancillary response file. The $0.7-8{\rm~keV}$ SXT spectrum is consistent with a power law photon index of $1.8\pm0.2$ modified by absorption column $N_H=(1.3\pm0.8)\times10^{21}{\rm~cm^{-2}}$ which is in excess of the Galactic column, $N_{\rm  H,MW}=9\times10^{19}$~cm$^{-2}$. The observed flux is $5.1\times10^{-12}{\rm~erg~cm^{-2}~s^{-1}}$ in the 0.7--8 keV  band.
We extracted the source spectrum and light curve using a circular region of radius $15^{\prime}$ centered at the source position. The poor spatial resolution of the SXT spreads the source photons almost over the entire detector area, thus leaving no source-free regions for background spectral extraction. Therefore, we used a background spectrum that was generated by the POC from a large number of blank-sky observations.  We used the redistribution matrix file available at the POC, and an updated ancillary response file. We grouped the spectral data to a minimum of 20 counts per bin, and analyzed using the spectral fitting package XSPEC version 12.12.0 \cite{xspec}. We fitted the  $0.7-8{\rm~keV}$ SXT spectrum with a power-law model modified by the Galactic and host galaxy absorption i.e., \texttt{ tbabs $\times$ ztbabs $\times$ zashift (powerlaw)} in the XSPEC terminology. We fixed the Galactic column at $N_{H,MW} = 9\times 10^{19}{\rm~cm^{-2}}$, obtained from the HEASARC column-density calculator\footnote{\url{https://heasarc.gsfc.nasa.gov/cgi-bin/Tools/w3nh/w3nh.pl}} \cite{heasarcnh}. We also fixed the redshift at $z=1.193$. This model resulted in an acceptable fit ($\chi^2 = 208.7$ for 231 degrees of freedom) with $\Gamma = 1.63_{-0.14}^{+0.15}$, the host galaxy absorption column of $2.9_{-2.7}^{+3.2}\times 10^{21}{\rm~cm^{-2}}$, and the absorption-corrected $0.7-8{\rm~keV}$ flux of $4.3\times 10^{-12}{\rm~erg~s^{-1}}~cm^{-2}$.

\subsubsection{\nicer}\label{supsec:nicer}
\nicer started high-cadence monitoring (multiple visits per day) of \target on 2022-02-16 19:07:03 (UTC) or MJD 59626.80, roughly 5 days after optical discovery. The resultant dataset comprises of several hundred snapshots {\bf, i.e., Good Time Intervals (GTIs),} whose exposures varied  between a few hundred to roughly 1200 seconds. In this work, we report data taken prior to MJD~59697 (28 April 2022), i.e., from the first 76 days after optical discovery.

% and defer analysis of the subsequent \nicer\ observations to future work.

%\section{No contaminating sources within \nicer's Field of View}
We started \nicer data analysis by downloading the raw, unfiltered ({\it uf}) data from the HEASARC public archive \footnote{\url{https://heasarc.gsfc.nasa.gov/cgi-bin/W3Browse/w3browse.pl}}. We reprocessed the data using the standard procedures outlined on the \nicer data analysis webpages (\url{https://heasarc.gsfc.nasa.gov/docs/nicer/analysis\_threads/}). We follow the data reduction steps outlined in \cite{2018cowpasham}.

\begin{figure}[htbp!]
    \centering
    \includegraphics[width = \columnwidth]{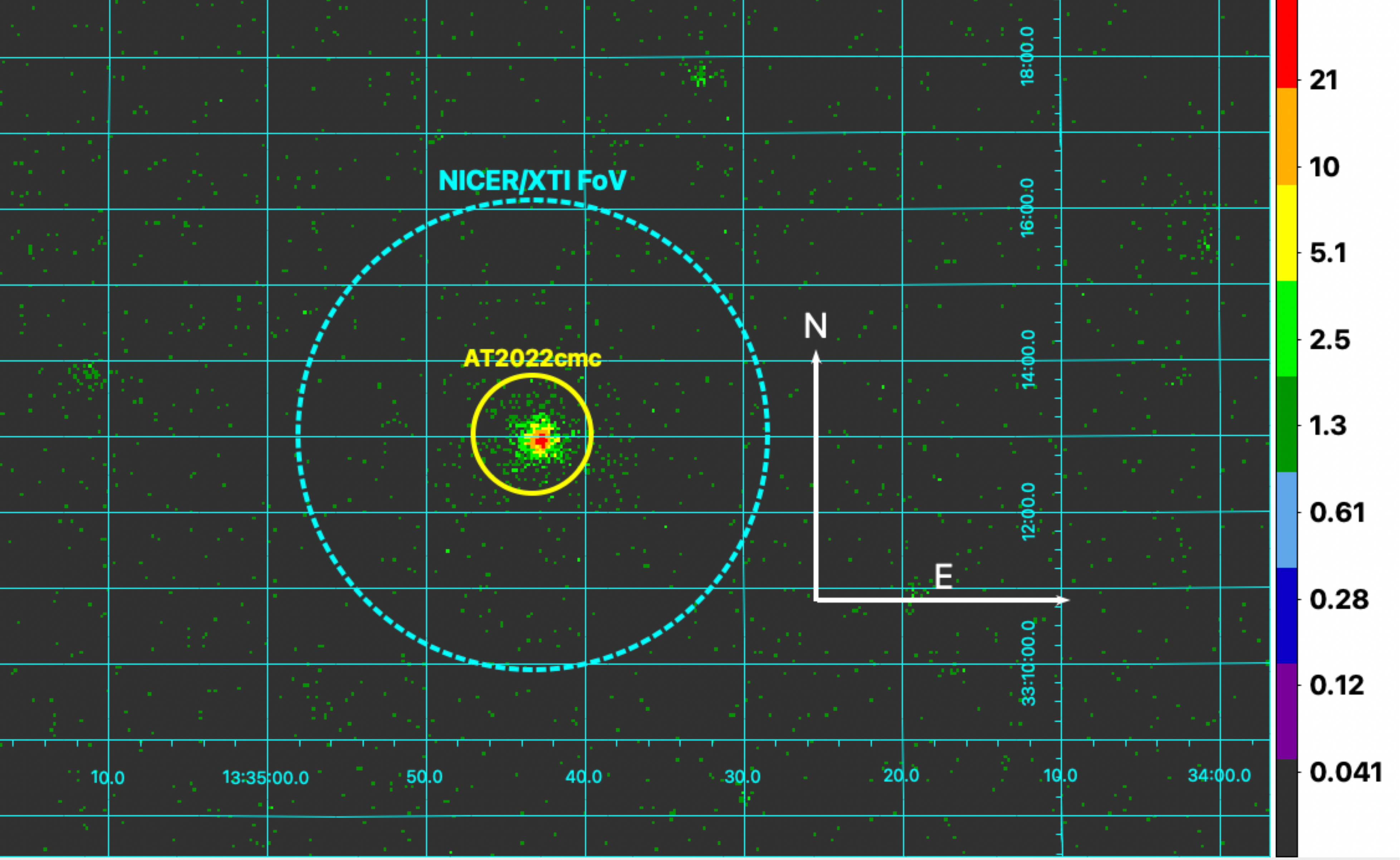}
    \caption{{\bf {\it Neil Gehrels Swift} XRT 0.3-8 keV image of \nicer's FoV}. The yellow  circle with a radius of 47$^{\prime\prime}$ and is centered on \target's radio coordinates of 13:34:43.2, +33:13:00.6 (J2000.0 epoch). The outer/dashed cyan circle shows {\it NICER}/XTI's approximate field of view of 3.1$^{\prime}$ radius. There are no contaminating sources within {\it NICER}'s FoV.  The north and east arrows are each 200$^{\prime\prime}$ long. The colourbar shows the number of X-ray counts.  \label{fig:xrtimage}}
\end{figure}

\nicer is a non-imaging instrument with a field of view (FoV) area of roughly 30 arcmin$^{2}$ (radius of 3.1$^{\prime}$). To test for the presence of potential contaminating sources in  \nicer's field of view, we extract a 0.3-8 keV X-ray image using \swift/XRT observations of the field (Extended Data Figure \ref{fig:xrtimage}). We find that % (obsIDs: 00015023001, 004, 005, 006, 009, 010, 011, 012). 
%It is evident from  Fig. \ref{fig:xrtimage} that 
\target is the only source within \nicer's FoV, implying that the flux from \target\ dominates the \nicer\ light curve at all times.

We investigate the X-ray spectral evolution of \target\ by extracting time-resolved spectra from the \nicer\ data taken between MJD~59626 and 59642 at $\approx0.5$~day intervals (\ref{tab: xraydata}). Spectral analysis from data beyond MJD 59642, i.e., where \target's flux is reduced and comparable to the \nicer background, will be published in a separate work. The main steps we follow are described below. 

\begin{enumerate}
    \item First, we extract the combined unfiltered but calibrated (ufa) and cleaned (cl) event files using the start and the end times of all GTIs within a given epoch.
    \item Then, we use the 3c50 background model \cite{Remillard2022} on these combined ufa and cl files to estimate the average background and source spectra. All the detectors marked as ``hot'' at least once in any of the individual GTIs are excluded. ''hot'' detectors are those affected by optical light loading (see \cite{2018cowpasham} for more description). A detector is tagged as ``hot'' if its 0.0-0.2 keV raw count rate is more than 4$\sigma$ above the median of all active (typically 52) \nicer detectors.
    \item Using the tools {\tt nicerarf} and {\tt nicerrmf} we extract an arf and rmf for each epoch.
    \item Then, we group the spectra using the optimal binning criterion described by \cite{Kaastra2016} also ensuring that each bin have at least 25 counts. We implemented this using the {\tt ftool} {\tt ftgrouppha} with {\it grouptype = optmin} and {\it groupscale = 25}.
    
\end{enumerate}

%\section{\nicer time-resolved spectral analysis}

\begin{figure}[htbp!]
    \centering
    \includegraphics[width = 0.75\columnwidth]{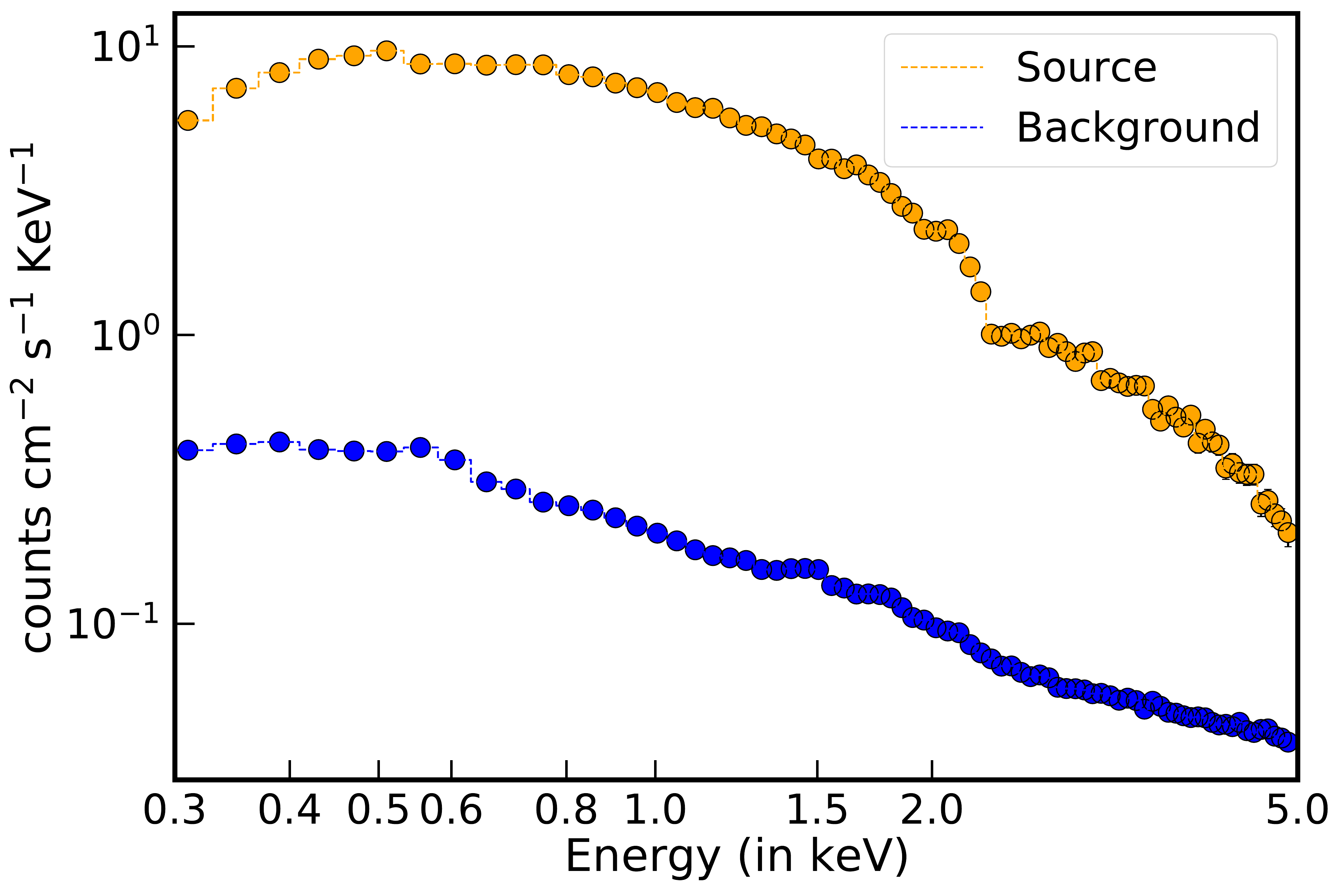}
    \caption{{\bf A sample \nicer X-ray spectrum}. The orange and the blue data represent the source and the estimated background spectra, respectively. This particular dataset is from the E0 epoch of the Extended Data Table \ref{tab: xraydata}. The 1$\sigma$ uncertainties are smaller than the data points. \label{fig:nicerspec}}
\end{figure}

\begin{figure}[htbp!]
\centering
\includegraphics[width=0.9\columnwidth]{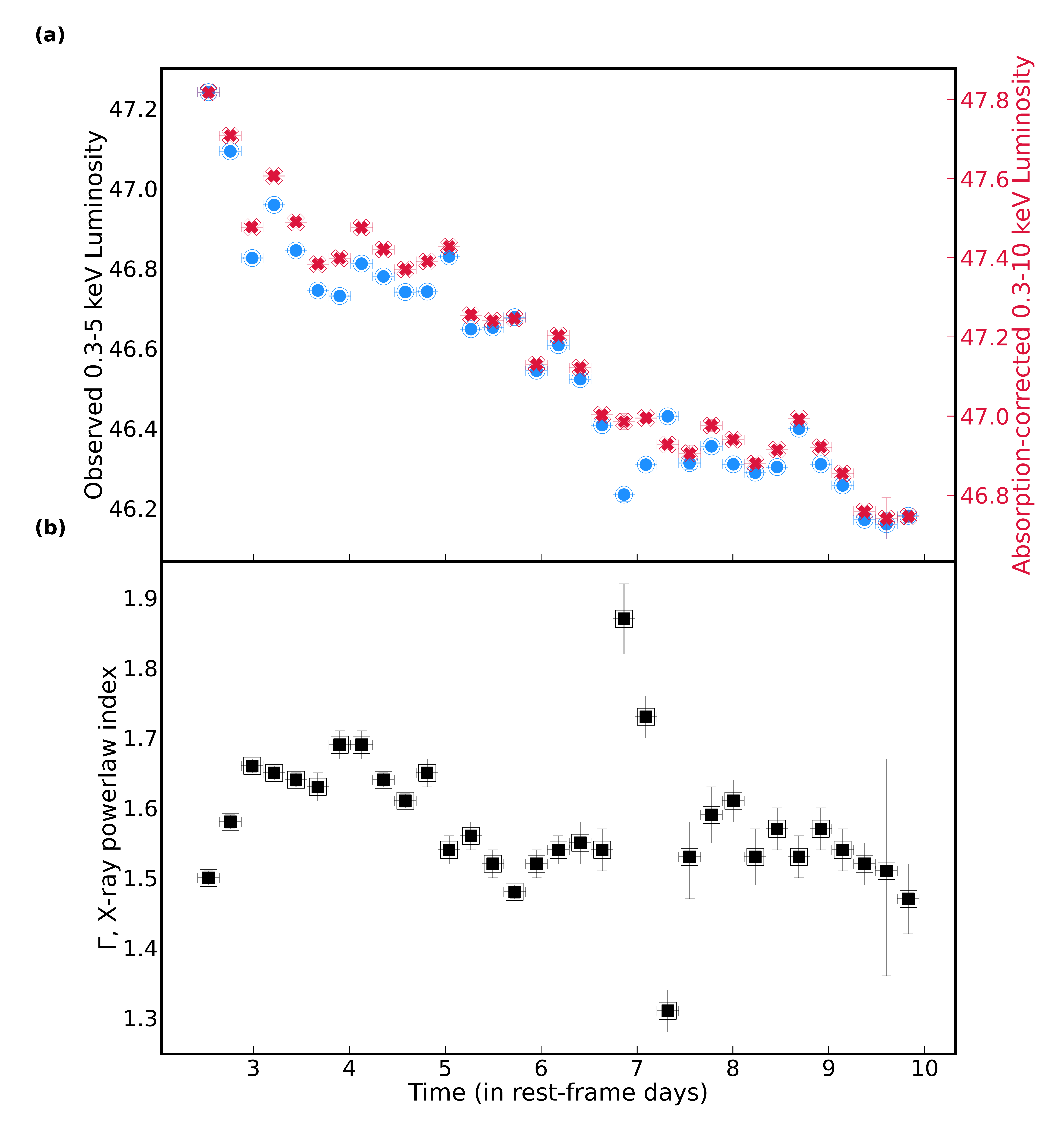}
\caption{ {\bf \target's X-ray luminosity and energy spectral slope evolution}. (a) Logarithm of the observed 0.3-5 keV (filled blue circles; left y-axis) and the absorption-corrected 0.3-10 keV luminosities (filled red crosses; right y-axis) in units of ergs~s$^{-1}$. The errorbars on the luminosities are much smaller than the size of the data points. (b) Evolution of the best-fit power-law index with time. The abrupt changes in index around day 7 (rest-frame) coincide with a hard X-ray (2--5 keV) flare that happened during epoch E21 (the data point with best-fit photon index of $\sim$1.3; see Extended Data Table \ref{tab: xraydata}). The neutral Hydrogen column of the host was tied across all epochs and the best-fit value is (9.7$\pm$0.3)$\times$10$^{21}$ cm$^{-2}$. All the errorbars represent 1$\sigma$ uncertainties. The individual \nicer spectra are posted at to a public repository at  \url{https://doi.org/10.5281/zenodo.6870587}.  }\label{fig:xrayevol}
\end{figure}

\begin{figure}[htbp!]
    \centering
    \includegraphics[width = \columnwidth]{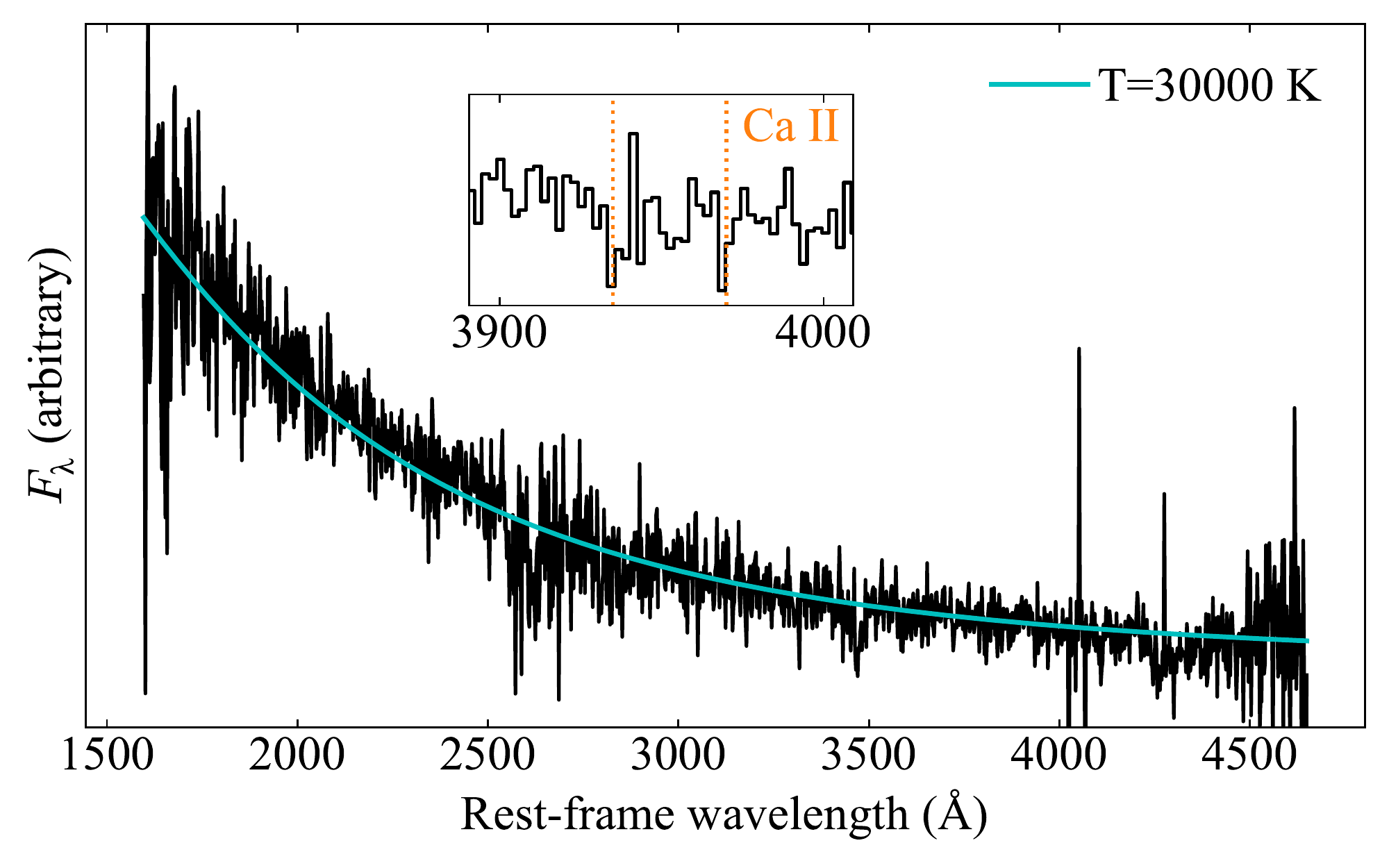}
    \caption{{\bf VLT/X-shooter spectrum of \target, obtained at $\approx15$~days after discovery.} The featureless blue continuum can be modelled with a blackbody with $T\approx 30,000$\,K (solid blue line), consistent with the optical bump in the broad-band SED from day 25-27 (Figure \ref{fig:SEDs}). The inset shows a zoom in on the region with CaII absorption lines identified by \cite{cmcredshift2}.  \label{fig:opticalspectrumvlt}}
\end{figure}

%dividing the \nicer data between MJD 59626 and 59642$\footnote{Spectral analysis from data beyond MJD 59642, i.e., where AT2022cmc's flux is reduced and comparable to the \nicer background, will be published in a separate work.}$ into several epochs of roughly 0.5 d in the observer frame and extracting their spectra (Table \ref{tab: xraydata}). 
We model the resulting time-resolved spectra in the 0.3-5.0 keV bandpass, the energy range in which the source was above the background using a \texttt{ tbabs $\times$ ztbabs $\times$ zashift (clumin*power-law)} model in {\it PyXspec}, a {\tt Python} implementation\footnote{\url{https://heasarc.gsfc.nasa.gov/xanadu/xspec/python/html/index.html}} of {\it XSPEC} \cite{xspec}. We fix the Milky Way column to $N_{\rm  H,MW}=9\times10^{19}$~cm$^{-2}$, estimated from the HEASARC nH calculator\footnote{\url{https://heasarc.gsfc.nasa.gov/cgi-bin/Tools/w3nh/w3nh.pl}} \cite{heasarcnh}. We tied the host galaxy neutral Hydrogen column to be the same across all the spectra and incorporated an additional 1\% systematic uncertainty while fitting the data\footnote{\url{https://heasarc.gsfc.nasa.gov/docs/nicer/analysis\_threads/cal-recommend/}}. The cosmological parameters were set in {\it XSPEC} to the values mentioned above. We set the {\it Emin} and the {\it Emax} parameters of {\it clumin} to 0.3 and 10.0, respectively. This allows us to compute the k-corrected, unabsorbed 0.3-10 keV luminosities at various epochs. A sample \nicer X-ray spectrum is shown in the Extended Data Figure \ref{fig:nicerspec}. We also tried a thermal model which resulted in strong systematic residuals throughout the X-ray bandpass considered and hence we did not consider it any further.

%As per the \nicer analysis guide  (\url{https://heasarc.gsfc.nasa.gov/docs/nicer/analysis\_threads/cal-recommend/}) we added an additional 1\% systematic uncertainty while modeling. The spectra were modeled using {\it PyXspec}, a {\tt Python} implementation of {\it XSPEC} (\url{https://heasarc.gsfc.nasa.gov/xanadu/xspec/python/html/index.html}).

The above modeling resulted in a total $\chi^{2}$/degrees of freedom (dof) of 2135.3/1956. The reduced $\chi^{2}$ values are close to unity in all expect during epoch E21 in which systematic residuals below 1 keV and above 5 keV are clearly present. This epoch coincides with a hard (2-5 keV) X-ray flare. Multiple such flares are evident between MJD 59637 and 59697. One such flare is also captured by \swift (see panel (d) of Figure \ref{fig:fig1}). We defer the spectro-timing analysis of these flares to a future work. 

% Adding a thermal component improved the fit from a $\chi^{2}$ of 125.5 to 58 with a loss of 2 dof. The origin of this thermal component is unclear but we speculate that this may be the inner disk of the black hole revealed during a jet wobbling episode similar to that claimed in Sw~J1644+57 \cite{swj1644mad:2014MNRAS.437.2744T}. The evolution of the the power-law index, host galaxy column, observed and absorbed-corrected luminosities are shown in Fig. \ref{fig:xrayevol}. 

% The apparent changes in $\Gamma$ could be due to a changing X-ray spectral break energy \cite{yaoxraybreak} with an increase in $\Gamma$ indicating a decrease in break energy and vice versa. But given \nicer's limited bandpass, we cannot test this hypothesis more robustly.

Following \cite{Remillard2022} we set \nicer's sensitivity limit to a conservative value of 0.3-5 keV count rate of 0.2 counts/sec (normalized to 50 \nicer detectors). In other words, any particular time segment in which the background-subtracted 0.3-5 keV countrate is less than 0.2 cps is treated as an upper limit of 7.4$\times$10$^{45}$ erg s$^{-1}$. This upper limit corresponds to k-corrected 0.3-10 keV absorption-corrected luminosity of 1.2$\times$10$^{46}$ erg~s$^{-1}$ for a source at a redshift of 1.193 (see panel (a) of Figure \ref{fig:fig1}).

\subsubsection{\swift/X-Ray Telescope(XRT)}\label{data:xrt}
\swift was not operational during the optical detection of \target and the satellite resumed pointed operations on 17 February 2022 \cite{swiftresumes}. \swift\ began monitoring \target on MJD 59633 (23 February 2022) and was observed under the ID of 00015023. The source was observed once a day between MJD 59633 and 59638 and once every few days after MJD 59638. {\bf In this work, we used data until MJD 59703, i.e., observation IDs 00015023001 through 00015023035}. We started our data analysis by downloading the raw, level-1 data from the HEASARC public archive and reprocessed them using the standard HEASoft tool {\tt xrtpipeline}. Here, we only consider the data taken in the Photon Counting (PC) mode. We only used events with grades between 0 and 12 in the energy range of 0.3 and 5 keV to match \nicer's bandpass. We extracted the source and background counts using a circular aperture of $47^{\prime\prime}$ and an annulus with an inner and outer radii of $80^{\prime\prime}$ and $200^{\prime\prime}$, respectively. XRT count rates were extracted on a per obsID basis and these values have been provided as a supplementary file named ''xrt\_0.3\_5.0keV.dat''.

To convert \swift/XRT count rates to fluxes we extracted an average energy spectrum by combining all the XRT exposures. We fit the 0.3-5.0 keV spectra with a power law model, modified by \target's host galaxy neutral Hydrogen column and MilkyWay, same as the model used for \nicer data above. Because the signal-to-noise of the \swift XRT spectrum is low, the host galaxy Hydrogen column was fixed at 9.8$\times$10$^{20}$ cm$^{-2}$ as derived from \nicer fits. We left the power law photon index free which yielded a best-fit value of 1.45$\pm$0.06. This value is consistent with \nicer spectral fits. From this fit we estimated the observed 0.3-5 keV flux and a count rate-to-flux scaling factor of $3.6\times10^{-11}\,{\rm erg}\, {\rm cm}^{-2}\,{\rm counts}^{-1}$ to covert from 0.3-5 keV background-subtracted XRT count rate to observed flux in the 0.3-5 keV band (Figure \ref{fig:multiband_lcs}). The uncertainties on the count rates, and consequently, the scaled fluxes were computed using the formulae for small number statistics described in \cite{gehrels1986}.

\subsubsection{GRB and TDE Comparison Data}\label{data:XrayComparison}
In order to compare the X-ray light curve of \target\ with other relativistic transients, we compile a sample of X-ray light curves of the three known relativistic TDEs, together with the  bright GRBs from \cite{Gompertz18}. 
For the GRBs in our comparison sample, we download the 0.3--10\,keV count-rate light curves from the UK \emph{Swift} Science Data Centre (UKSSDC) \cite{Evans07,Evans2009} and correct them for absorption using the ratio of time-averaged unabsorbed flux to time-averaged observed flux per burst, provided in the UKSSDC catalog\footnote{\url{https://www.swift.ac.uk/xrt\_live\_cat/}}. We k-correct the light curves to rest-frame 0.3--10\,keV luminosity following \cite{Bloom01}, assuming a power-law spectrum with photon index given by the time-averaged photon-counting mode photon index from the UKSSDC catalog.

We extract X-ray light curves of the three relativistic TDEs using the UKSSDC XRT products builder\footnote{\url{https://www.swift.ac.uk/user\_objects/}} \cite{Evans07,Evans2009}. We use a time bin size of one day. We convert the 0.3--10\,keV count rate light curves to unabsorbed flux using the counts-to-flux ratio of the time-averaged spectral fits, and k-correct them to rest frame 0.3--10\,keV as described above. The X-ray spectral indices for Sw~J1644+57 and Sw~J2058+0516 were variable between 1.2-1.8 \cite{seifina17}. This range is similar to \target (see the Extended Data Table \ref{tab: xraydata}). Here we used the following fiducial values: Sw~J1644+57: cts:flux $= 9.32\times10^{-11}$\,erg\,cm$^{-2}$\,ct$^{-1}$, photon index $= 1.58\pm0.01$; Sw~J1112.2-8238: cts:flux $= 6.13\times10^{-11}$\,erg\,cm$^{-2}$\,ct$^{-1}$, photon index $= 1.35\pm 0.08$; Sw~J2058.4+0516: cts:flux $= 5.36\times10^{-11}$\,erg\,cm$^{-2}$\,ct$^{-1}$, photon index $= 1.55\pm0.08$. We plot these light curves, together with the GRB X-ray light curves extracted above, in Figure~\ref{fig:fig1}.

\subsection{UV/Optical Observations}
\subsubsection{Zwicky Transient Facility}
\target was discovered and reported by the Zwicky Transient Facility (ZTF; \cite{ztf}) and 
released as a transient candidate ZTF22aaajecp
in the public stream to brokers and the Transient Name Server, with data available in Lasair\footnote{\url{https://lasair.roe.ac.uk/object/ZTF22aaajecp}}
\cite{Lasair:Smith}. 
We performed point spread function (PSF) photometry on all publicly available ZTF data using the ZTF forced-photometry service \cite{Masci_2019} in $g$- and $r$-band. We report our photometry, corrected for Galactic extinction of $A_{\rm V}=0.0348$~mag \cite{Schlafly_2011} and converted to flux density in mJy, in Extended Data Table~\ref{tab: allphot}.

\subsubsection{ATLAS}
The Asteroid Terrestrial-impact Last Alert System (ATLAS; \cite{2018PASP..130f4505T}) is a $4\times0.5$ meter telescope system, providing all-sky nightly cadence at typical limiting magnitudes of $\sim 19.5$ in cyan ($g+r$) and orange ($r+i$) filters. The data are processed in real time and the transients are identified by the ATLAS Transient Science Server \cite{2020PASP..132h5002S}. %\target was not flagged by the server since it did not result in a minimum of three co-spatial $5\sigma$ detections on any night. 
We stacked individual nightly exposures and used the ATLAS forced photometry server \cite{2021TNSAN...7....1S} to obtain the light curves of \target\ in both filters. Photometry was produced with standard PSF fitting techniques on the
difference images and 
we initially reported the fast declining optical flux in \cite{atlasLC:Fulton}. 

\subsubsection{Follow-up optical imaging}
Followup of \target~ was conducted as part of the ``advanced” extended Public ESO Spectroscopic Survey of Transient Objects (ePESSTO+) \cite{smartt2015} using the EFOSC2 imaging spectrograph at the ESO New Technology Telescope to obtain images in $g$, $r$ and $i$ bands. Images were reduced using the custom PESSTO pipeline (\url{https://github.com/svalenti/pessto}), and the PSF photometry was measured without template subtraction using \emph{photometry-sans-frustration}; an interactive python wrapper utilising the Astropy and Photutils packages \cite{mnichollpsf}. Aperture photometry was applied to the few images in which the target PSF was slightly elongated, otherwise the magnitudes were derived from PSF-fitting. All photometry has been calibrated against Pan-STARRS field stars.
%\textcolour{red}{NEED TO ADD ePESSTO+ here (Michael/Stephen?)}
% DONE 

%\subsubsection{EFOSC}
%EFOSC DISCUSSION TO GO HERE
% I'm not sure we need all the subsections...
% Stephen - I agree, we just need one on optical imaging. I have rearranged. 

%\subsubsection{Other optical imaging}
\target was also followed up in $r$, $i$, $z$ and $w$ bands with the 1.8 meter PanSTARRS2 (PS2) telescope in Hawaii \cite{Chambers2016}.  PS2 operates in survey mode, searching for near-Earth objects but the survey can be interrupted for photometry of specific targets. PS2 is equipped with a 1.4 Gigapixel camera with a pixel scale of $0.26^{\prime\prime}$. The images were processed with the Image Processing Pipeline (IPP; \cite{2020ApJS..251....5M}) and difference imaging was performed using the PS1 Science Consortium (PS1SC; \cite{Chambers2016}) 3$\pi$ survey data as reference. PSF photometry was used to compute instrumental magnitudes, and zero-points were calculated from PS1 reference stars in the field.

\target was also observed as part of the Kinder (kilonova finder) survey \cite{2021TNSAN..92....1C} in $g$, $r$, and $i$ bands with the 0.4m-SLT at Lulin Observatory, Taiwan. The images were reduced using a standard {\sc IRAF} routine with bias, dark and flat calibrations. We used the AUTOmated Photometry Of Transients (AutoPhOT) pipeline \cite{2022arXiv220102635B} to perform PSF photometry and calibrate against SDSS field stars \cite{2022TNSAN..39....1C}. We used the Lulin one-meter telescope (LOT) for deeper imaging in $g$, $r$, $i$ and $z$ bands over four nights spanning 13.4--16.2~days after discovery. The images were also reduced using the standard CCD processing techniques in {\sc IRAF}. We performed aperture photometry calibrated against SDSS field stars. 
In a combined stack of the images from the LOT, \target was clearly detected in $g$, $r$ and $i$ bands, with magnitudes $21.76\pm0.14$, $21.71\pm0.18$ and $21.93\pm0.31$ mag, respectively and undetected in $z$ band with an upper limit of $>20.69$ mag. We list the photometry from our individual observations in the Extended Data Table~\ref{tab: allphot}.

We compile additional optical photometry from the GCN circulars \cite{gcn31592,gcn31593,gcn31594,gcn31597,gcn31625,gcn31647,gcn31652,gcn31729,gcn31798,gcn31805,gcn31846} and correct for extinction. These are also included in the Extended Data Table~\ref{tab: allphot}. 
%In this work we also included all the data on this source published on the GCN and the Astronomy Telegram channels as of 25 April 2022.

\subsubsection{\swift/UVOT}
We perform photometry on \swift/UVOT \cite{Roming_2005} observations of \target\ with the \textit{uvotsource} task in HEAsoft package v6.29 using a 5$^{\prime\prime}$\ aperture on the source position. Another region of 40$^{\prime\prime}$~located at a nearby position was used to estimate the background emission. Because the host galaxy is not detected in the GALEX \cite{Bianchi_11} coadded UV images and  \target's UVOT detections are $\sim$ 2 mag brighter then host upper limits (see ``Constraints on host luminosity''), we did not attempted any type of host subtraction.
%We correct the derived photometry for Galactic foreground extinction \cite{Schlafly_2011}. For non-detections we report the 3$\sigma$ upper-limits.

\subsubsection{\textit{AstroSat}/UVIT}
The \textit{AstroSat} Ultra-Violet Imaging Telescope (UVIT\cite{2017AJ....154..128T,2020AJ....159..158T}) onboard {\it AstroSat}\cite{2014SPIE.9144E..1SS} also observed the source, simultaneous with the SXT, with its Far Ultra-violet (FUV) channel using the F148W($\lambda_{mean}=1481{\text{\AA}}$; $\Delta\lambda=500{\text{\AA}}$) and F154W ($\lambda_{mean}=1541{\text{\AA}}$; $\Delta\lambda=380{\text{\AA}}$) filters for exposures of $6024{\text{s}}$ and $9674{\text{s}}$, respectively. We processed the level1 data using the CCDLAB pipeline\cite{2017PASP..129k5002P} and constructed broadband images. We extracted source counts using a circular aperture of radius 10$^{\prime\prime}$\ centered at the source position. We also extracted background counts from nearby source-free regions, and corrected for the background contribution. We then converted the net count rates to the flux densities using the flux conversion factors provided in \cite{2017AJ....154..128T,2020AJ....159..158T}. We do not detect the source, and obtain 3-$\sigma$ flux upper limits of $4.7\times10^{-17}$ erg~cm$^{-2}$~s$^{-1}$~\text{\AA}$^{-1}$ (F154W) and $6.4\times10^{-17}$  erg~cm$^{-2}$~s$^{-1}$~\text{\AA}$^{-1}$ (F148W).

\subsubsection{Optical spectroscopy}
\label{data:optspec}
We observed \target with the X-shooter spectrograph \cite{Vernet2011} on the European Southern Observatory's Very Large Telescope (VLT) on 27 February 2022. Data were obtained in on-slit nodding mode using the 1.0$^{\prime\prime}$, 0.9$^{\prime\prime}$, and 0.9$^{\prime\prime}$ slits in the UVB, VIS and NIR arms respectively, with a spectral resolution of $\approx 1$\,\AA\ in the optical. We reduced the data following standard procedures \cite{Selsing2019a}. We first removed cosmic-rays with the tool  {\tt astroscrappy}\footnote{\href{https://github.com/astropy/astroscrappy}{https://github.com/astropy/astroscrappy}}, which is based on cosmic-ray removal algorithm by \cite{vanDokkum2001a}. Afterwards, we processed the data with the X-shooter pipeline v3.3.5 and the ESO workflow engine ESOReflex \cite{Goldoni2006a, Modigliani2010a}. We reduced the UVB and VIS-arm data in stare mode to boost the signal to noise by a factor of $\sqrt{2}$ compared to the standard nodding mode reduction. We co-added the individual rectified and wavelength- and flux-calibrated two-dimensional spectra, followed by extraction of the one-dimensional spectra of the each arm in an statistically optimal way using tools developed by J. Selsing\footnote{\href{https://github.com/jselsing/XSGRB\_reduction\_scripts}{https://github.com/jselsing/XSGRB\_reduction\_scripts}}.% using tools by J. Selsing.
Finally, we converted the wavelength calibration of all spectra to vacuum wavelengths and corrected the wavelength scale for barycentric motion. We stitched the spectra from the UVB and VIS arms by averaging in the overlap regions. We reduced the NIR data reduced in nodding mode to ensure a good sky-line subtraction. We do not detect a trace of the target in the NIR arm and thus do not discuss the NIR data further.

The extracted spectrum consists of a steep and largely featureless blue continuum, which we rebin by 5 pixels to increase the signal to noise (Extended Data Figure~\ref{fig:opticalspectrumvlt}). At the reported redshift $z=1.193$, there is a hint of absorption features at wavelengths consistent with the Ca\,II H\&K lines. The apparent absorption at $\sim2600$~\AA{} is not a real feature, but rather a low-sensitivity, noisy region close to the edge of the UVB arm. The spectrum (covering rest-frame $\sim1500-4500$\,\AA) can be well fit by a blackbody with $T\approx 30,000$\,K, though a power law with $F_\nu \propto \nu^{0.6}$ also provides a satisfactory fit. The thermal model is preferred due to its consistency with the optical bump in the broad-band SED (Figure~\ref{fig:SEDs}). This value is consistent with the  measurement of $\sim$2.3$\times$10$^{4}$ K from the optical/UV SED, after accounting for the synchrotron contribution and the measurement uncertainty of $\sim$10\% on the value inferred from the VLT spectrum. This inferred temperature is similar to other optical TDEs \cite{vanVelzen2021}.

\subsubsection{Constraints on host luminosity}\label{supsec:host}
In order to put upper limits on the luminosity of the host galaxy, we created deep reference images in $w, i, z$ bands by stacking PanSTARRS1 and PanSTARRS2 images of the field containing \target. These images were obtained during routine survey operations over a period spanning June 2010 to January 2022.
The $w$-band is a wide filter ($3900-8500$ \AA) with an effective wavelength $\lambda_{\mathrm{eff}} \approx 6000$ \AA, and can thus be treated as $r$-band. The effective exposure time for the co-added reference stacks is 2475~s, 13700~s, 16260~s, in $w, i, z$ bands respectively. The host galaxy of \target is not visible in any of these stacks, with upper limits of $w > 23.85$, $i > 23.05$ and $z > 22.89$ mag (see Extended Data Figure \ref{fig:legacyimage}).

The deepest observer-frame limit ($r-$band) corresponds to rest-frame absolute AB magnitude of $M_{2740}>-19.9$, with a simple $k$-correction of $2.5\log(1+z)$ and the observer frame central wavelength converted to rest-frame (approximately 2740\AA), with only a Milky Way reddening correction applied to the observer frame flux. The redder bands similarly correspond to $M_{3430}>-20.7$ and 
$M_{3950}>-20.8$. We performed a similar analyses on GALEX \cite{Bianchi_11} $NUV$ ($\lambda_{\mathrm{eff}} \approx 2300$ \AA) and $FUV$ ($\lambda_{\mathrm{eff}} \approx 1535$ \AA) filters data by stacking all images that contains the position of \target. No underlying host emission is detected in any of stacked images, and the 3$\sigma$ upper limits are $NUV > 22.6$ and $FUV > 22.5$ mag. 

\subsection{Radio}
\subsubsection{VLA}
We observed \target\ on 2022 February 27 ($\approx15$ d after discovery) with NSF's Karl G. Jansky Very Large Array (VLA) under program 20B-377 (PI: Alexander). The observations were taken when the array was in its most extended A configuration. We used the C, X, Ku, K, and Ka band receivers with the 3-bit digital samplers to obtain nearly continuous frequency coverage from $4-37$ GHz. We used 3C286 for bandpass and flux density calibration. We used J1329+3154 for complex gain calibration at K and Ka bands, and 3C286 otherwise.  We reduced and imaged the data using standard procedures in the Common Astronomy Software Applications (\textsc{CASA}) v5.6.1-8 \cite{McMullin2007}. We detect a bright unresolved point source at all frequencies, enabling us to split the data into 2 GHz bandwidth segments for photometry. The resulting SED is shown in Figure~\ref{fig:SEDs}.

\subsubsection{Arcminute Microkelvin Imager - Large Array}
The Arcminute Microkelvin Imager -- Large Array (AMI-LA) is a radio interferometer consisting of eight 12.8\,metre dishes with baselines from 18 to 110\,metres, located in Cambridge, UK \cite{Zwart2008}. AMI-LA observes at 15.5\,GHz with a bandwidth of 5\,GHz divided into 4096 channels \cite{Hickish2018}. We observed \target\ with AMI-LA beginning 14.7\,days after discovery \cite{cmcdiscovery:Igor}. We reduced the AMI-LA observations using a custom pipeline \textsc{reduce\_dc} \cite{Anderson2018}. The pipeline averages the data down to 8 channels, performs flagging for radio frequency interference and antenna shadowing. We used 3C286 for both amplitude and complex gain calibration. We performed additional flagging, imaging and deconvolution in \textsc{casa} (Version 4.7.0).
%using the \textit{clean} task \cite{McMullin2007}. 
We combine the statistical uncertainty on the 15.5\,GHz flux densities with a 5\% systematic calibration uncertainty in quadrature. 
We detected an unresolved source with a flux density of 0.49$\pm$0.03\,mJy in the first epoch \cite{gcn31667}, and initiated subsequent observations at near-daily cadence. We present the full 15.5\,GHz light curve in Figure~\ref{fig:multiband_lcs} and list the flux density measurements in Extended Data Table~\ref{tab: allphot}. 
We compile additional radio measurements of \target\ reported online in GCN circulars and Astronomer's Telegrams \cite{gcn31592,gcn31665,atel15269} together in Extended Data Table~\ref{tab: allphot}.

\subsubsection{EVN sub-milliarcsecond position}\label{methods:evn}
We used the European Very Long Baseline Interferometry (VLBI) Network (EVN) to observe \target on 2022 March 22--23 (18:08--02:11 UTC), under project code RM017A (PI: Miller-Jones), making use of the real-time eVLBI mode. We observed in dual-polarization mode, at a central frequency of 4.927\,GHz. Our array consisted of 15 stations, with ten standard EVN stations (Jodrell Bank Mk II, Effelsberg, Hartebeesthoek, the 16-m dish at Irbene, Medicina, Noto, the 85$^{\prime}$ dish at Onsala, the 65-m dish at Tianma, Torun, and Yebes) that observed with a bandwidth of 256\,MHz, and five stations from the eMERLIN array (Knockin, Darnhall, Pickmere, Defford, and Cambridge), which observed with a reduced bandwidth of 64\,MHz. 

We processed the data through the EVN pipeline to derive the a priori amplitude calibration and bandpass corrections, and conducted further processing with the Astronomical Image Processing System (AIPS, version 31DEC19 \cite{Greisen2003}). We phase referenced the data on \target to the nearby (1.66$^{\circ}$ away) calibrator source J1329+3154, with an assumed position of (J2000) 13:29:52.864912, +31:54:11.05446. We detected \target as an unresolved point source with a significance of $6.4\sigma$, at a position of (J2000) 13:34:43.201308(6), +33:13:00.6506(2). The quoted uncertainties (denoted in parentheses for the last significant digit) are purely statistical, with potential systematic errors (e.g.\ from uncorrected tropospheric delay or clock errors) estimated to be at the level of $\sim0.07$\,mas.

\section{Shortest X-ray variability timescale} \label{xrayvariability}
Manual inspection of the 0.3-5 keV background-subtracted \nicer light curve of \target (provided as a supplementary file) reveals multiple instances of a variation in the observed count rate by $>50$\% within a span of a few hundred seconds. To quantify the  variability timescale, we extracted an average power density spectrum (PDS) using uninterrupted exposures that were each 950~s long\footnote{Increasing the accumulation time to 1024~s exposures yields fewer  samples (13, compared to 29) and only results in a marginal gain in low frequency information from 1/950~Hz to 1/1024~Hz).} within the first month of discovery, i.e., data acquired before MJD 59642 (rapid flaring activity observed at later times will be considered in a separate work). To ensure minimal impact from background fluctuations, we only considered exposures that were above the background, i.e., background-subtracted 0.3-5 keV count rates greater than 0.2 counts/s (normalized to 50 \nicer detectors), close to the nominal limit described by \cite{Remillard2022}. In addition to the standard filters described in ``$\gamma$-ray and X-rays/\nicer'' we impose a filter to remove exposures where the observed mean 15-18 keV count rate is beyond two standard deviations of the median 15-18 keV rate measured across all exposures. This is an extra-cautionary step to minimize the effect of background particle flaring which is important for variability studies. This gives a total of 29 time series with a cumulative exposure of 27.55 ks (950$\times$29). We compute a Leahy-normalized (\cite{leahy}; mean Poisson noise level of 2) average power density spectrum (PDS) sampled at 1/8 seconds from these time series (Extended Data Figure~\ref{fig:figpds}). We find that the PDS is consistent with the Poisson noise level of 2 at high frequencies ($\gtrsim10^{-2}$~Hz); however, the PDS  starts to rise above the noise level at $\lesssim2\times$10$^{-3}$ Hz, and the lowest-frequency bin at 1/950 s clearly well-above the noise level. This suggests that \target has systematic X-ray variability on timescales at least as short as $\sim$1000~s in observer frame. 

\section{Arguments against a GRB afterglow}\label{supsec:notgrb}
\begin{figure}[htbp!]
    \centering
    \includegraphics[width = \columnwidth]{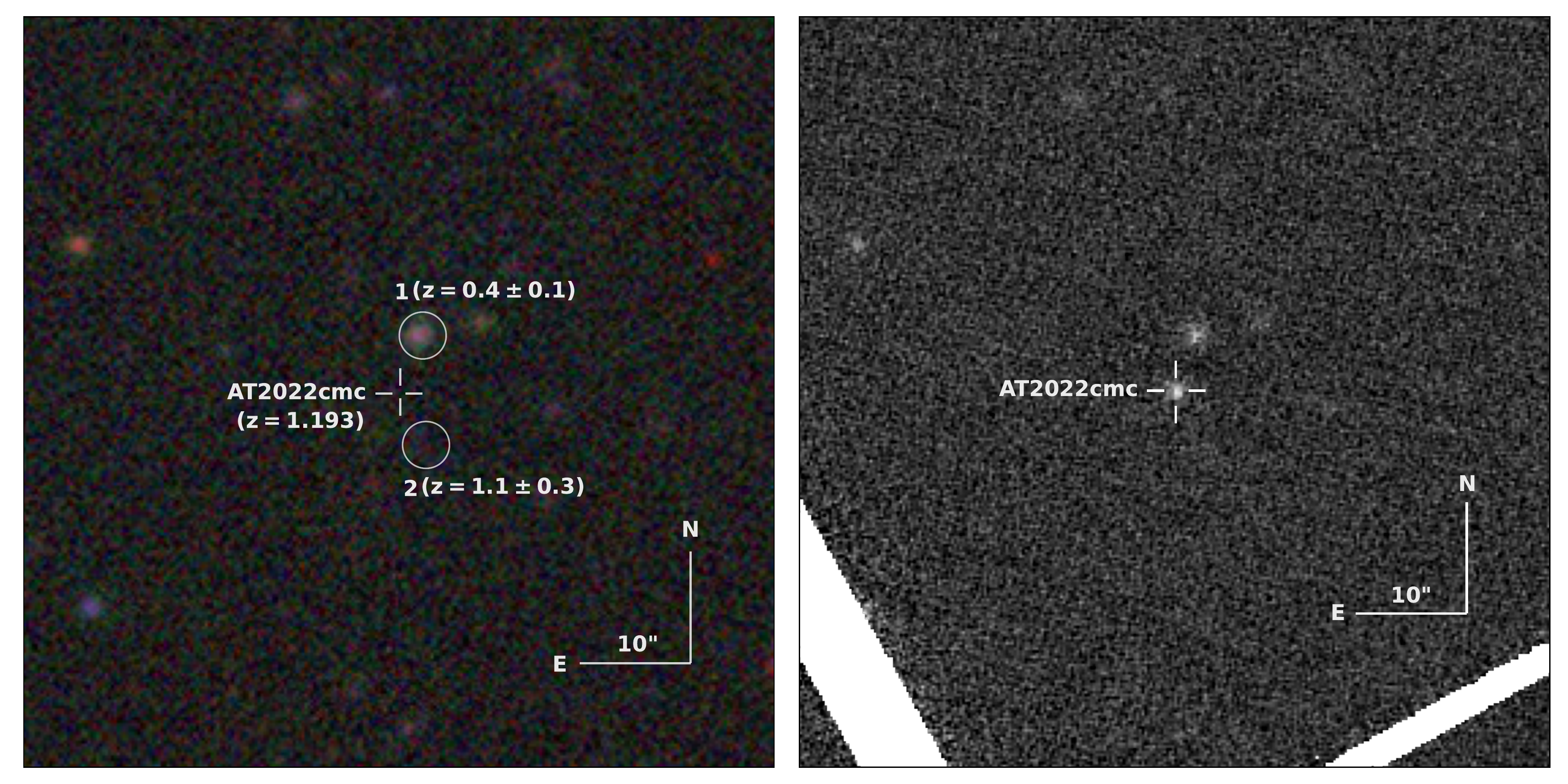}
    \caption{{\bf Average X-ray (0.3-5 keV) power density spectrum of \target}. The frequency resolution and the Nyquist frequency are 1/950 Hz and 1/8 Hz, respectively. This power spectrum is an average of 29 individual PDS. The dashed, red curve is the best-fit power-law model. Systematic variability on timescales of $\sim$1000 s (lowest frequency bin) is evident. All the frequencies and hence the timescales are as measured in observer frame. The errorbars represent 1$\sigma$ uncertainties. \label{fig:figpds}}
\end{figure}

\begin{figure}[htbp!]
    \centering
    \includegraphics[width = \columnwidth]{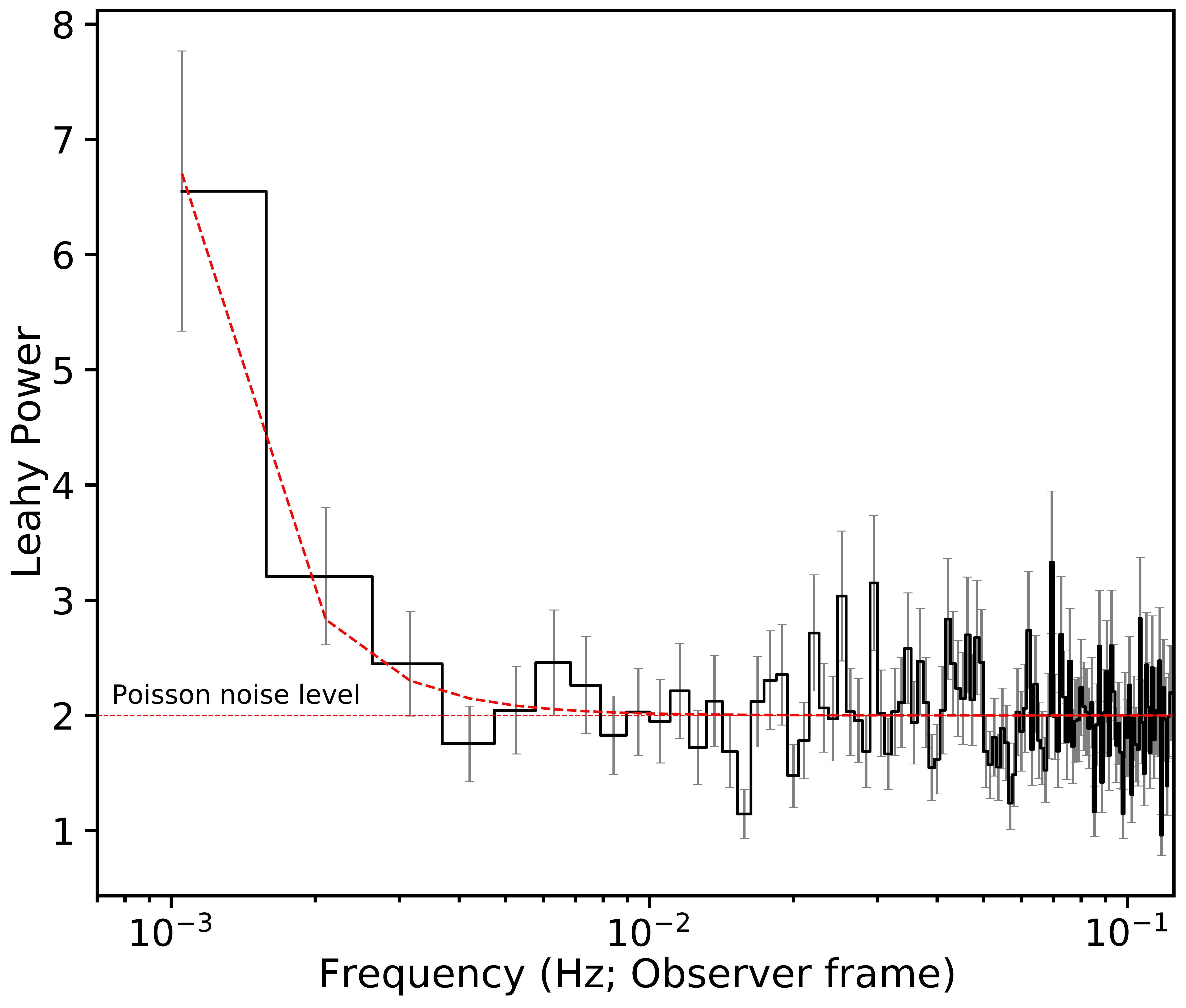}
    \caption{{\bf Pre and post-outburst optical images of \target.} Left panel: A colour composite image of the field prior to the outburst, made using data from the Legacy Imaging Surveys \cite{2019AJ....157..168D} using g, r and z filters. There is no emission at the location of \target (cross). Nearby catalogued objects with their photometric redshifts are shown (circles). Right panel: A PS2 $w$-band image of \target post outburst. The size of both image cutouts is $1.1^{\prime} \times1.1^{\prime}$. North and the East arrows are each 10$^{\prime\prime}$.
    %The x- and y- units represent the camera pixels. 
    \label{fig:legacyimage}}
\end{figure}

A potential association with the \textit{Fermi} Gamma Ray Burst (GRB) 220211A \cite{GCN31570} was ruled out following a more precise localization of that GRB \cite{Ridnaia22}. Nevertheless, the early optical evolution resembled an off-axis gamma-ray burst (GRB). 
Long GRBs occur as a result of the core-collapse of massive stars (e.g., \cite{Woosley93, LongGRBSNe:McFadyen1999,GRbreview:KumarandBing}). Their emission comes in two phases: prompt emission, which consists of high-energy $\gamma$-rays generated within the ultra-relativistic jet that is launched following collapse \cite{Blandford76,Paczynski86}, and the afterglow, which is produced by shocks as the jet is decelerated in the environment surrounding the burst \cite{Rees92,Sari98}. High-cadence \nicer\ and \swift/XRT monitoring observations have shown that \target has been consistently brighter than even the most luminous known GRB afterglows by more than a factor of 10 (see panel (a) of Figure~\ref{fig:fig1}). The most striking difference between \target\ and GRB afterglows is the persistence of rapid X-ray variability (e.g., Figure~\ref{fig:fig1} panels (a)-(d), and see Extended Data Figure~\ref{fig:figpds}). The \nicer\ observations reveal short ($\approx2.4$~hrs observer frame, corresponding to $\approx1$~hr in the source rest frame) flares with increases in the count rate by factors of 2--10 that remain detectable until at least $\approx40$~days after discovery. This variability requires that the X-ray emitting region be smaller than $R = 2\Gamma_{\rm j}^2 c \delta t$ $\approx 10^{-4}\Gamma_{\rm j}^2$~parsec (where $\Gamma_{\rm j}$ is the bulk Lorentz factor of the jet). In contrast, the expected tangential radius of a GRB afterglow at a similar time is $\approx0.5$~pc for typical parameters \cite{gs02} and $\Gamma_{\rm j}\lesssim2$. Continued central engine activity, which operates at much smaller radii ($\sim 10^{13}$~cm, e.g. \cite{Pe'er15}) may produce rapid variability \cite{Kobayashi97}, but even the longest GRBs (the so-called `ultra-long' class; \cite{Levan14}) do not show signs of central engine activity beyond a day after trigger (e.g. \cite{Zhang14}). On the other hand, X-ray variability on timescales of tens of minutes has been inferred for the relativistic TDEs, Sw~J1644+57 \cite{Saxton2012} and Sw~J2058+05 \cite{Pasham2015}. These properties strongly favour a non-GRB origin.

\section{Multi-wavelength SED modeling}
\subsection{Preliminary Considerations}
\label{preliminary_considerations}
\begin{figure}[htbp!]
\begin{center}
\includegraphics[width=\textwidth, angle=0]{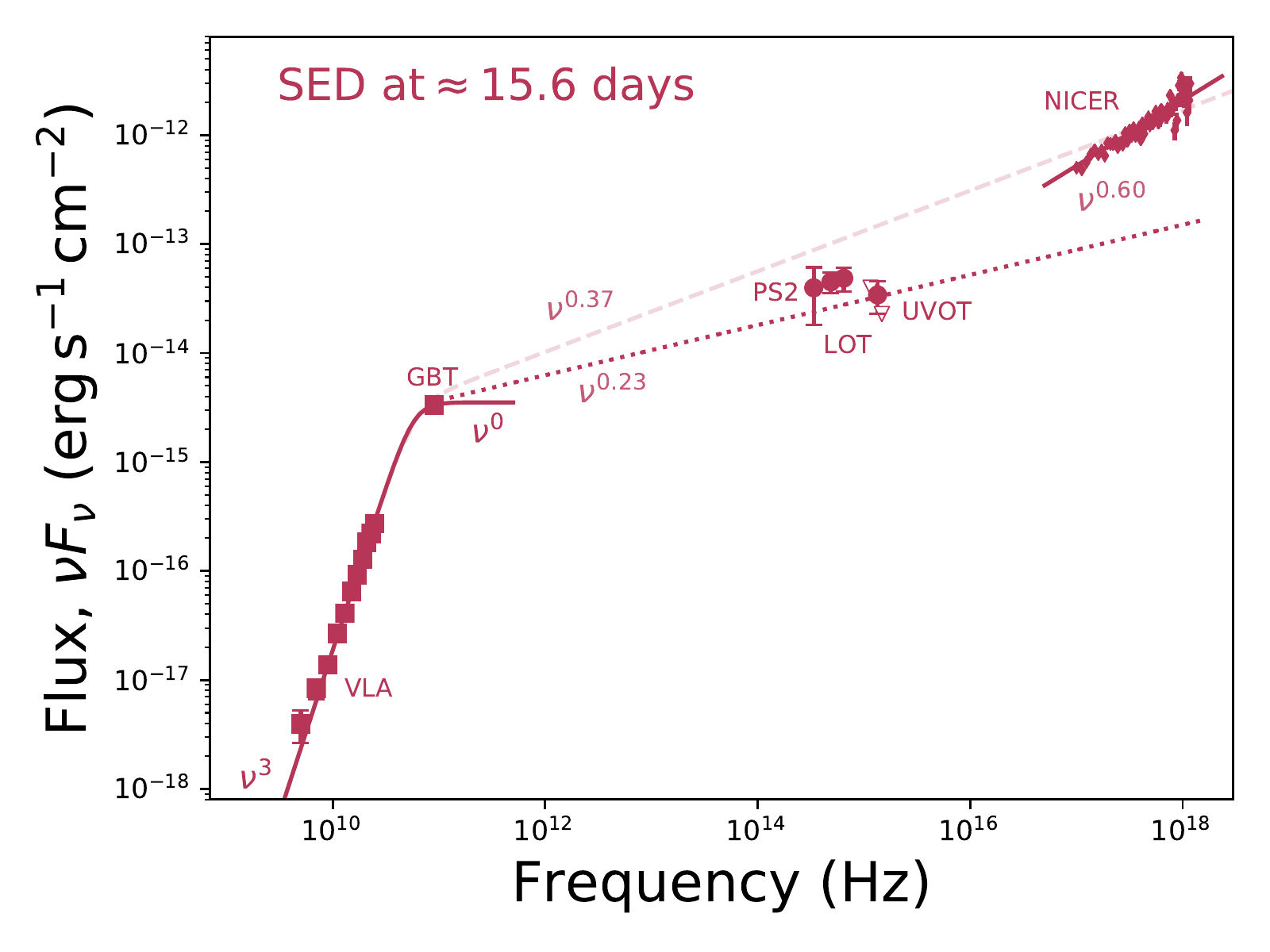}
\end{center}
\caption{{\bf Spectral energy distribution of \target\ at $\approx15.6$~days after discovery.} Data at radio (VLA), mm-band (GBT), UV/optical (\swift/UVOT, ZTF, PanSTARRS) and X-ray frequencies (\nicer), demonstrate that the SED at this time cannot be explained as  a single synchrotron spectrum. The SED at $\lesssim25$~GHz is optically thick ($\nu F_{\nu}\propto\nu^3$), with a spectral break near $\approx90$~GHz. The spectral index from the GBT observation at $\approx90$~GHz to the \nicer\ band is $\nu F_{\nu}\propto \nu^{0.37}$, which (i) is significantly shallower than the observed \nicer\ spectral index ($\nu F_{\nu}\propto\nu^{0.57}$) and (ii) significantly over-predicts the UV flux at this time. All the errorbars represent 1$\sigma$ uncertainties. }
\label{fig:SED_analytical}
\end{figure}

The full multi-wavelength (radio to X-ray) spectral energy distribution of \target\ cannot be simply explained by synchrotron emission. To see this, we consider the SED at $\approx15.6$~days after discovery (Extended Data Figure~\ref{fig:SED_analytical}) at radio (VLA), mm-band (GBT), ultraviolet (\swift/UVOT) and X-ray frequencies (\nicer). The start and the end times of the GBT observation were MJD 59637.2868 and 59637.2928. We find that the spectral index from the GBT mm-band ($90$~GHz) observation to the center of the \nicer\ X-ray band is $\beta_{\rm mm-X}=-0.63\pm0.01$ (corresponding to $\nu F_\nu\propto \nu^{0.37}$). This is inconsistent with the observed hard \nicer\ spectrum, $\beta_{\rm X}=-0.40\pm0.02$ (corresponding to $\nu F_\nu\propto \nu^{0.60}$). Furthermore, the interpolation from the radio to the X-rays using the above spectral index over-predicts contemporaneous \swift/UVOT \textit{UM2}-band observations (when corrected for Galactic extinction) by a factor of $\approx4$. This is unlikely to be explained by UV variability, which appears to be $\lesssim20\%$ at this time. While extinction due to dust could suppress the UV flux, there is no evidence for significant dust extinction along the line of sight, as evidenced by the blue $z^\prime-g^\prime\approx-0.1$~mag colour as well as the blue optical spectrum at this time (Section~\ref{data:optspec}). The absence of significant extinction is further confirmed by the \hst\ \textit{F160W} and \textit{F606W} measurements at $\approx25.4$~days, which yield a spectral index of $\beta_{F606-F160}=0.34\pm0.08$. Thus, it is not possible to extend a single power-law spectrum from the radio to the X-rays without a mismatch between the required spectral index and the observed X-ray spectral index, and without over-predicting the optical/UV flux, indicating that the radio and X-ray flux arise from distinct emission components at this time. 

Furthermore, the optical SED at this time appears to peak in $\approx g$-band, with a spectral index $\beta_{g-um2}=-1.5\pm0.5$. This declining spectral index cannot connect with observed X-ray flux, as the spectral index between the optical and X-rays at this time is much harder, $\beta_{\rm opt-X}\approx-0.2$. This suggests that the optical and X-ray emission at this time also arises from separate emission components. This is further confirmed by the very different temporal evolution in the X-rays ($\alpha_{\rm X}\approx-2.2$ and optical ($\alpha_{\rm r^\prime}\approx-0.3$) at $\approx10$--40~days post-discovery. 

The radio SED at $\lesssim25$~GHz is optically thick ($\beta\approx2$), whereas the spectral index between the flux density measured with the VLA 24.5~GHz and with the GBT at 90~GHz is $\beta_{\rm K-mm}=-0.96\pm0.06$, indicating a spectral break is present near the GBT frequency. A simple broken power-law fit to the radio-mm SED at this time with the post-break index fixed at $\beta\approx-1$ yields a break frequency of $\nu_{\rm pk}=(57.5\pm0.1)$~GHz and a spectral peak flux density of $F_{\nu,\rm pk}=(4.1\pm0.1)$~mJy at 15.6~days. Identifying this as the peak of a synchrotron SED, a simple energy equipartition argument suggests a minimum kinetic energy of $E_{\rm K,iso}\approx10^{50}$~erg and radius of $R_{\rm eq}\approx10^{16}$~cm for this component \cite{bdnp13}. In the next section, we relax the assumption of equipartition and perform a full model fit with a physical model including SSC emission in the X-rays and a black body component in the optical. 

\subsection{Model setup}
For our model fits, we create three SEDs of \target by combining the data taken on days 15-17, 25-27, and 41-46, as these epochs have the best multi-wavelength coverage. In each of these SED epochs we only had single measurements in the optical, the UV filters and the various radio bands. However, multiple \nicer/X-ray exposures were present. These were merged to extract combined spectra using the procedure outlined in section \ref{supsec:nicer}. We fit each SED with a simple homogeneous single zone model, similar to those used for blazars, e.g. \cite{Ghisellini09,Boettcher13,Tavecchio16}. In this model, a power-law energy distribution of electrons with number density $n_{e}$, energy index $p$, and minimum and maximum Lorentz factors $\gamma_{\rm min}$ and $\gamma_{\rm max}$, is injected in a spherical region of radius $R$, threaded with a magnetic field $B$ and moving with a bulk Lorentz factor, $\Gamma_{\rm j}$ with respect to the observer at viewing angle, $\theta$. The quantities $B$, $n_{e}$ and $R$ are calculated in the emitting region co-moving frame. 
We test two different model setups in order to probe which radiative mechanisms are responsible for the high energy emission. In the simplest case (which we call model 1), we consider synchrotron and SSC exclusively. In the second case, we test a simple external inverse Compton model (model 2 from now on), in which the seed photons are provided by the optical black body component \footnote{Unlike \cite{Burrows2011}, we can not test whether the seed photons originate in the accretion disk, as this component is not detected in any of the SEDs we model and is therefore entirely unconstrained.}. 

Modelling the UV/optical emission as, e.g., a disk wind is very complex and beyond the scope of this work \cite{crumley}. Given the thermal appearance of the UV/optical SED, we make the simplifying assumption that this is black body emission originating in a thin shell at a radius $R_{\rm bb}=(L_{\rm bb}/4\pi\sigma_{\rm sb}T_{\rm bb}^{4})^{1/2}$ (in analogy with how blazar jet models typically treat the torus around the AGN, e.g. \cite{Ghisellini09}), and derive $L_{\rm bb}$ and $T_{\rm bb}$ from the temperature and normalization of the thermal component as we run the fit. In order to estimate the relative contribution of EC and SSC we need to calculate the energy density in the co-moving frame of the jet. For this, we need to assume an opening angle $\phi$ to convert the radius of emitting region $R$ to a distance from the central engine. For simplicity, we take $\phi = 1/\Gamma_{\rm j}$ and estimate the distance from the black hole to be $d=R/\phi = \Gamma_{\rm j} R$. Finally, we calculate the black body energy density $U_{\rm bb}$ as follows. For $d<R_{\rm bb}$, the emitting region in the jet is moving towards the black body (in which case EC is expected to contribute meaningfully to the SED) and we have simply $U_{\rm bb}=\Gamma_{\rm j}^{2}L_{\rm bb}/(4\pi R_{bb}^{2}c)$. For $d\geq R_{\rm bb}$, we account self consistently (following the prescription in \cite{Lucchini21} for an AGN torus) for the de-boosting of the photons, as the jet emitting region is moving away, rather than towards, the optical-emitting region. This choice of jet opening angle means that the efficiency of EC is maximized with respect to SSC. This is because maximizing the jet opening angle (by setting $\phi = 1/\Gamma_{\rm j}$) minimizes the distance $d$ from the black hole for a given radius $R$, which in turn makes it more likely that the optical photons will be Doppler-boosted in the frame of the jet. We note that for AGN jets, VLBI surveys find typical values of $\phi \approx 0.1-0.2\Gamma_{\rm j}$ \cite{Pushkarev09}. This smaller opening angle would push the emitting region farther away from the black body, reducing the efficiency of EC. The cyclo-synchrotron and inverse Compton emission are calculated using the \texttt{Kariba} libraries from the \texttt{BHJet} publicly available model \cite{Lucchini21}.

\begin{figure}[htbp!]
\begin{center}
\includegraphics[width=\textwidth, angle=0]{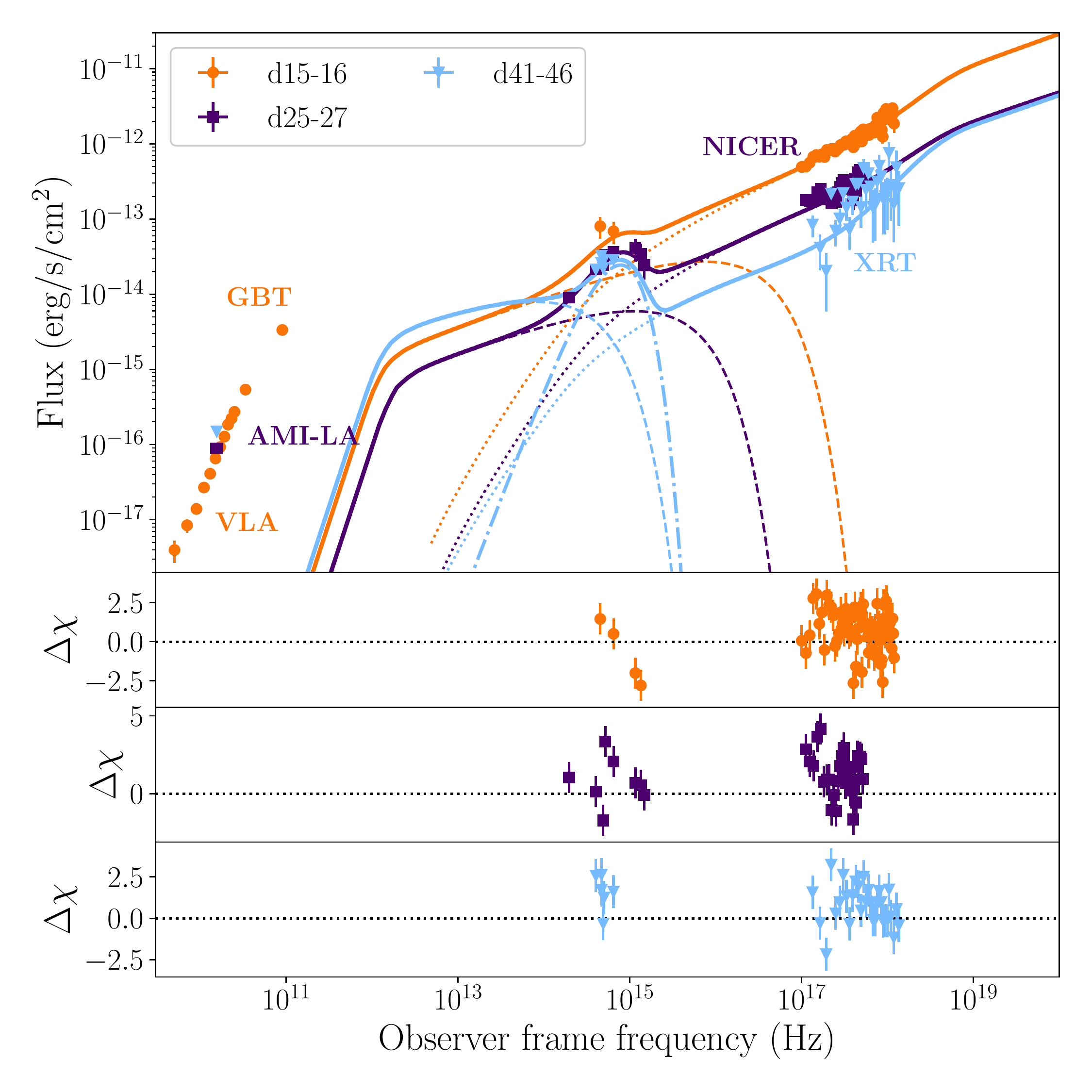}

\end{center}
\caption{{\bf Contour plots for the best-fitting parameters of model 1.} For clarity, we only show the 2d posterior distributions of parameters that are degenerate with each other.}
\label{fig:Posteriors}
\end{figure}

\begin{figure}[htbp!]
\begin{center}
\includegraphics[width=\textwidth, angle=0]{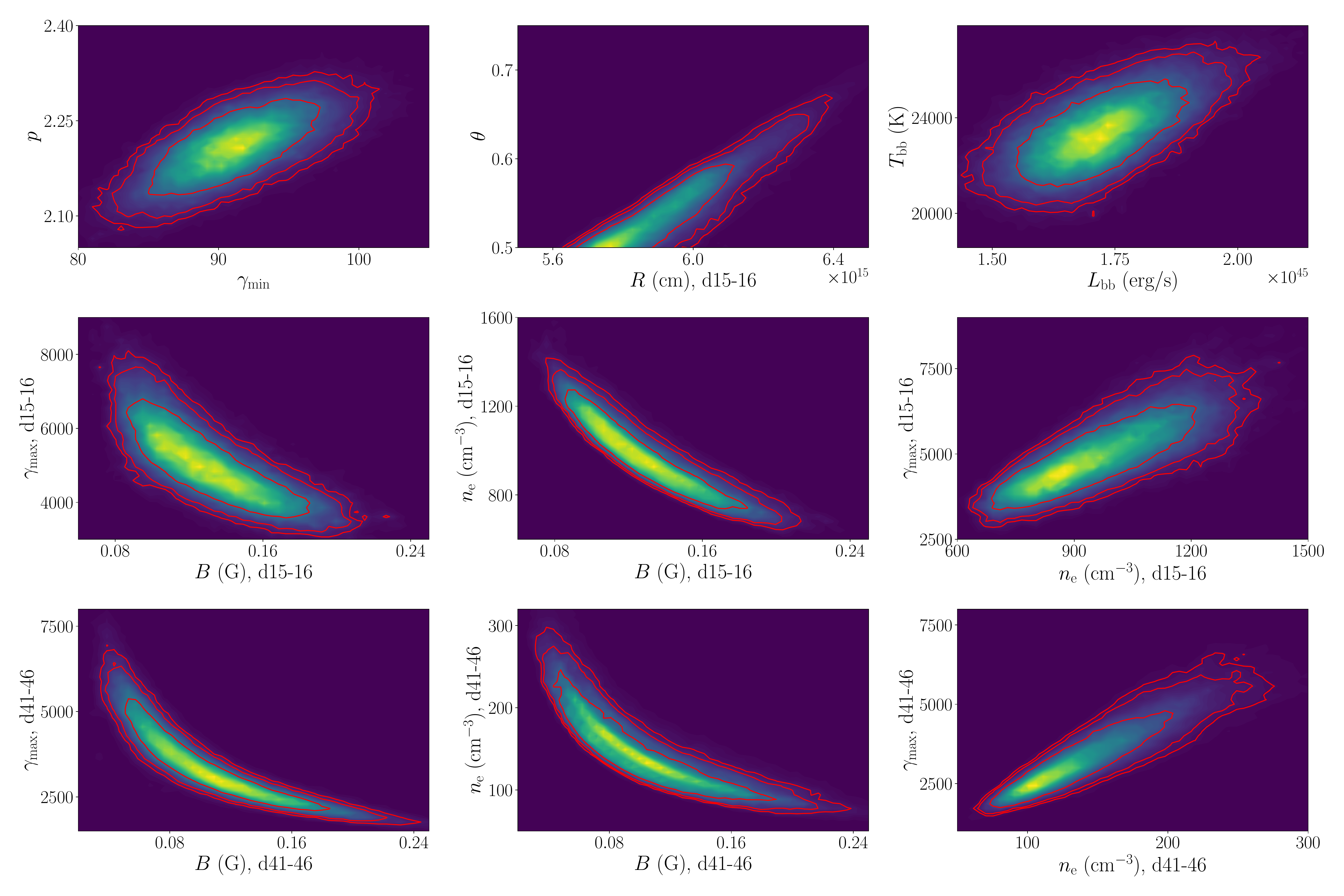}
\end{center}
\caption{{\bf Best fitting External inverse Compton (EC) model}. The EC model requires a jet that under-predicts the radio flux. Furthermore, EC produces too little soft X-ray flux, and as in model 1 the emission at these frequencies is dominated by SSC. All the errorbars represent 1$\sigma$ uncertainties. }
\label{fig:SED_EC}
\end{figure}

We import the data and model into the spectral fitting package \texttt{ISIS}, version 1.6.2-51 \cite{isis} and jointly fit the SEDs at the three epochs. We tie the minimum Lorentz factor $\gamma_{\rm min}$, the particle distribution slope $p$, the bulk Lorentz factor $\Gamma_{\rm j}$ and the viewing angle $\theta$ across all epochs (meaning the parameters are free during the fit, but forced to be identical for each SED) and jointly fit all three SEDs, aiming to simplify the parameter space as much as possible. To obtain a starting guess for the model parameters, we perform an uncertainty-weighted least-squares fit using the $\chi^{2}$ statistic with the \texttt{subplex} minimization algorithm. We then explore the parameter space via Markov Chain Monte Carlo (MCMC) with \texttt{emcee} \cite{Foreman-Mackey2013} using 50 walkers for each free parameter (for a total of 900 walkers). We run the MCMC for 15000 steps and discard the first 6000 as ``burn-in''. We report the median and $1\sigma$ credible intervals (corresponding to $68\%$ of the probability mass around the median) on each parameter, as well as additional derived quantities of interest, in Extended Data  Table~\ref{tab: SED_fits}. We present the model corresponding to the median values of the parameters in Figures~\ref{fig:SEDs} and Extended Data Figure \ref{fig:SED_EC} for models 1 and 2, respectively. We also show the 2d posterior distributions of the best-fitting parameters (for model 1) that exhibit some degeneracy in Extended Data Figure~\ref{fig:Posteriors}.

\subsection{Modelling results}

In the case of model 1, we find that all the model parameters are well constrained by the data with minimal degeneracy, as is typical of single-zone models (e.g.\cite{Tavecchio98,RadiativeBook}). The constraints are weaker for model 2, but the model parameters remain fairly well determined. This behaviour can be understood as follows. The SED samples 7 observable quantities: the synchrotron self-absorption frequency $\nu_{t}$ (set by the multiple radio points on the day 15-16 SED), the synchrotron luminosities in the optically thin and thick regimes $L_{\rm s, thin}$ and $L_{\rm s, thick}$ (constrained by the radio and optical data), the inverse Compton luminosity $L_{\rm ssc}$ (set by the NICER data), the X-ray photon index, the synchrotron scale frequency $\nu_{s}$, and the inverse Compton scale frequency $\nu_{c}$. The free parameters in the model affect each observable quantity differently, and as a result it is possible to relate one to the other. For example, the bolometric synchrotron luminosity scales as $L_{\rm s} \propto n_{e}R^{3}B^{2}\delta^{4}$, while the SSC bolometric luminosity scales as $L_{ssc}\propto n_e R^{3}\delta^{4}U_{\rm s}$, with $U_{\rm s}=L_{s}/4\pi R^{2}c\delta^{4}$. As a result, $L_{ssc}\propto n_{e}^{2}B^{2}R^{4}\delta^{4}$, so that $L_{ssc}/L_{s}\propto n_e R$: for a fixed synchrotron luminosity, the large X-ray luminosity observed with \nicer requires a large number density and/or a large emitting region. In similar fashion, $B$, $n_e$, $R$ and $\delta$ are further constrained by the dependency of $\nu_{t}$, $L_{\rm s, thick}$, $\nu_{s}$ and $\nu_{c}$ on the model parameters. The constraints on the remaining model parameters are more intuitive. The slope of the electron distribution $p$ is determined by the slope of the X-ray spectra, because (to first order) a power-law electron distribution produces a power-law SSC spectrum with spectral index, $\beta = (1-p)/2$. Finally, once $B$ and $\delta$ are determined, the minimum and maximum particle Lorentz factors $\gamma_{\rm min}$ and $\gamma_{\rm max}$ are constrained by requiring that the synchrotron spectrum fall between the radio and optical frequency, and that the low energy end of the SSC spectrum fall between UV and X-ray energies.

The main results of model 1 are as follows. First, we require the jet to be highly relativistic ($\Gamma_{\rm j} = 86^{+10}_{-9}$), viewed at a very small angle ($\theta \leq 1^{\circ}$) and very powerful ($\approx 10^{46-47}\,\rm{erg s^{-1}}$, depending on the epoch and jet matter content). For comparison, this power is near or at the Eddington luminosity of a $10^{8}M_{\odot}$ black hole (roughly the largest black hole mass for which a main sequence star can be tidally disrupted). Second, the size of the emitting region is $\approx 10^{15}-10^{16}\,\rm{cm}$, which is marginally consistent with the observed variability time-scale of $\approx1000$~s, thanks to the strong beaming ($\delta \approx 100$). Finally, all of our best-fitting models require the energy density of the electrons ($U_{e} = \langle\gamma\rangle n_e m_e c^{2}$, where $\langle\gamma\rangle$ is the average Lorentz factor of the radiating electrons) to be larger than that of the magnetic field ($U_b = B^{2}/8\pi$) by a factor $\approx 10^{2}$ (up to $10^5$ for days 25-27, although this number is likely driven by our choice of tying multiple parameters), implying that the bulk of the jet power is carried by the matter, rather than the magnetic field.

The picture is quite different in the case of model 2. First, this model requires a small emitting region radius ($R \approx 10^{14}$ cm) and jet Lorentz factor ($\Gamma_{\rm j}\approx 5$). This behavior occurs because if EC is to contribute meaningfully to the SED, the emission has to originate close enough to the black hole that $d\leq R_{\rm bb}$, so that the external photons are Doppler boosted in the jet co-moving frame. Invoking a smaller emitting region results in larger estimates for the magnetic field $B$ and electron number density $n_{\rm e}$. In turn, this causes the synchrotron self absorption frequency to move to $\approx 10^{12}$ Hz, well above where the observed break lies in the data, and suppressing the predicted radio flux as a result. Consequently, the EC model predicts negligible radio flux, and the radio emission in this model must originate in a separate region. Requiring not one but two individual, self-absorbing active regions in the jet means that this EC model would require significantly more fine-tuning than the SSC model. We account for the inability of the EC model to reproduce the observed radio flux by neglecting the radio data entirely in the final model 2 fits (not doing so causes the fit to either recover the model 1 fits, or produce fits with $\chi^{2}/d.o.f \approx 70$, rather than $\approx 2.3$ without the radio data). Neglecting the constraints provided by the self-absorbed synchrotron data also means that the best-fitting parameters for model 2 are less well determined. Additionally, for seed black body photons peaking at $\nu_{\rm bb}\approx10^{15}$ Hz, the EC component only begins to be important at a frequency $\nu_{\rm EC}\approx  \delta \Gamma_{\rm j}\gamma_{\rm min}^{2}\nu_{\rm bb}\approx 10^{18}$ Hz \cite{RadiativeBook}. This scaling causes the EC component to only produce bright hard X-ray and/or soft $\gamma$-ray emission, while under-predicting the soft X-ray flux. Instead, at frequencies $\leq 10^{18}\,\rm{Hz}$ the bulk of the flux is still produced through SSC, as in model 1. A similar behavior is also found when modelling the SEDs of powerful blazars \cite{Ghisellini09,Boettcher13,Hayashida15}, in which the X-ray emission typically originates through SSC, while the $\gamma$-ray emission is dominated by EC. Similarly to model 1, producing a large soft X-ray flux through SSC requires the jet to again be matter dominated, with $U_{\rm e}/U_{\rm b}\approx 100$. Finally, model 2 requires smaller jet powers, with $P_{\rm j}\approx 10^{45}$ erg s$^{-1}$. 

In summary, model 1 can satisfactorily fit the data at every epoch, although requiring a very highly beamed, matter-dominated jet. Model 2 on the other hand greatly under-predicts the radio data, which instead requires some fine-tuning in the form of a second self-absorbed emitting region further downstream. While in this case the beaming requirements are less severe, a large SSC contribution is still required to match the X-ray flux, resulting in a similarly matter-dominated jet to model 1. Due to all these considerations, we favour model 1 over model 2, with the caveat that our treatment of the EC process is fairly simplistic. Despite this caveat, the models presented here provide strong evidence that the emission of \target originates in a relativistic jet pointed towards Earth.

% \section{How far would an \target-like event be discovered with present and future X-ray facilities}

\section{Estimate of gravitational lens magnification by a foreground structure}\label{methods:lensing}
% Graham: I realise the following may be too long for the paper, but it gives you the background to the estimated magnification of ~5-10%. i hope it's useful. 
The high luminosity of AT~2022cmc motivates considering whether gravitational lensing by a foreground structure along the line of sight has magnified the flux that we detect. AT~2022cmc is located $5.6^{\prime\prime}$ from the galaxy SDSS\,J133443.05$+$331305.7, at a photometric redshift of $z=0.4\pm0.1$, and $3.7^\prime$ from the galaxy group WHL\,J133453.9$+$331004 at a spectroscopic redshift of $z=0.4$ \cite{Wen2015}. The optical luminosity of the group, and the sky location and colours of this galaxy are consistent with our line of sight to \target passing adjacent to a star-forming galaxy located in the infall region of ($R\simeq r_{200}$) of a galaxy group with a mass $M_{200}\simeq3\times10^{13}\,M_\odot$, where the mass estimate is obtained by combining the optical luminosity from \cite{Wen2015} with the mass-observable scaling relations from \cite{Mulroy2019}. To estimate lens magnification by the group, we assume an NFW density profile with concentration $c_{200}=5$, and adopt the formalism from \cite{Wright2000} to estimate a magnification of $\mu\simeq1.02$, i.e.\ just a $\simeq2$ per cent magnification of the flux. To estimate magnification by the galaxy, we compare its apparent magnitude in red pass-bands (i.e., relatively insensitive to any ongoing star formation) with a model for a passively evolving stellar population formed in a burst at a redshift of $z>2$. This yields an estimated luminosity relative to the luminosity function of cluster and group galaxies \cite{Lin2004} of $\simeq0.3L^\star$. Combining this estimate with the scaling relations between mass and luminosity commonly used to estimate galaxy masses in gravitational lens models (e.g., \cite{Richard2010}) we obtain a velocity dispersion estimate for the bulge of the galaxy of $\sigma\simeq120\,\rm km\,s^{-1}$. Then, adopting a singular isothermal sphere (SIS) model of the galaxy mass distribution, and using the standard expressions for the lensing properties of an SIS (e.g., \cite{Smith2022}), we derive an estimated Einstein radius of $\theta_{\rm E}\simeq0.25^{\prime\prime}$ and lens magnification of $\mu\simeq1.05$, based on the lens redshift of $z_{\rm L}=0.4$ and source redshift of $z_{\rm S}=1.193$. In summary, the lens magnification suffered by \target appears to be modest at $\mu\simeq1.05-1.1$, and cannot account for the high observed luminosity of the X-ray to radio counterpart.

%%%%%%%%%%%%%%%%%%%%%%%%%%%%%%%%%%%%%%%%%%%%%%%%%
%%%%%%%%%%%%%%%%%%%%%% Swift/XRT image %%%%%%%%%%%%%%%%%%%%%%%%%%%%%%%%%%%%%%%%%%%%%%%%%
\clearpage

%%%%%%%%%%%%%%%%%%%%%%%%%%%%%%%%%%%%%%%%%%%%%%%%%
%%%%%%%%%%%%%%%%%%%%%% Sample NICER spectrum %%%%%%%%%%%%%%%%%%%%%%%%%%%%%%%%%%%%%%%%%%%%%%%%%
\clearpage
%%%%%%%%%%%%%%%%%%%%%%%%%%%%%%%%%%%%%%%%%%%%%%%%%%%
%%%%%%%%%%%%%%%%%%%%%%%%%%%%%%%%%%%%%%%%%%%%%%%%%%%
%%%%%%%%%%%%%%%%%%%%%%%%%%%%%%%%%%%%%%%%%%%%%%%%%%%

%%%%%%%%%%%%%%%%%%%%%%%%%%%%%%%%%%%%%%%%%%%%%%%%%
%%%%%%%%%%%%%%%%%%%%%% NTT blue spectrum  %%%%%%%%%%%%%%%%%%%%%%%%%%%%%%%%%%%%%%%%%%%%%%%%%
%%%%%%%%%%%%%%%%%%%%%%%%%%%%%%%%%%%%%%%%%%%%%%%%%
%%%%%%%%%%%%%%%%%%%%%% NICER X-ray PDS %%%%%%%%%%%%%%%%%%%%%%%%%%%%%%%%%%%%%%%%%%%%%%%%%
\clearpage
%%%%%%%%%%%%%%%%%%%%%%%%%%%%%%%%%%%%%%%%%%%%%%%%%
%%%%%%%%%%%%%%%%%%%%%% legacy image  %%%%%%%%%%%%%%%%%%%%%%%%%%%%%%%%%%%%%%%%%%%%%%%%%
%%%%%%%%%%%%%%%%%%%%%%%%%%%%%%%%%%%%%%%%%%%%%%%%%%%
%SED for analytical arguments
\newpage
\vfill\eject
%%%%%%%%%%%%%%%%%%%%%%%%%%%%%%%%%%%%%%%%%%%%%%%%%%%
%%%%%%%%%%%%%%%%%%%%%%%%%%%%%%%%%%%%%%%%%%%%%%%%%%%
%model 1 posteriors
\newpage
\vfill\eject

%%%%%%%%%%%%%%%%%%%%%%%%%%%%%%%%%%%%%%%%%%%%%%%%%%%
%%%%%%%%%%%%%%%%%%%%%%%%%%%%%%%%%%%%%%%%%%%%%%%%%%%
%model 2 SED
\newpage
\vfill\eject

%%%%%%%%%%%%%%%%%%%%%%%%%%%%%%%%%%%%%%%%%%%%%%%%%%%

\ttabbox[\linewidth]{
\resizebox{\textwidth}{!}{\begin{tabular}{*{9}{l}}
\toprule
\toprule
\multicolumn{9}{c}{The first few entries of the multi-wavelength data presented in this work.  } \\
\bottomrule
{\bf Time} & {\bf Observatory} & {\bf Instrument} & {\bf Filter} & {\bf Frequency} & {\bf Flux} & {\bf Flux Error} & {\bf Detection?} & {\bf data}  \\
 {\bf (days)} &  &  & & {\bf (Hz)} & {\bf (mJy)} & {\bf (mJy)} & {\bf (1=Yes)} & {\bf source} \\
\midrule
$1.03\times10^{0}$ & ATLAS & NA & \textit{o} & $4.52\times10^{14}$ & $8.93\times10^{-2}$ & $8.62\times10^{-3}$ & 1 & This work \\
$1.05\times10^{0}$ & ZTF & NA & \textit{g'} & $6.46\times10^{14}$ & $5.93\times10^{-2}$ & $3.37\times10^{-3}$ & 1 & This work \\
$1.07\times10^{0}$ & ZTF & NA & \textit{r'} & $4.90\times10^{14}$ & $8.71\times10^{-2}$ & $3.27\times10^{-3}$ & 1 & This work \\
$2.07\times10^{0}$ & ATLAS & NA & \textit{o} & $4.52\times10^{14}$ & $5.05\times10^{-2}$ & $6.42\times10^{-3}$ & 1 & This work \\
. & . & . & . & . & . & . & . & . \\
. & . & . & . & . & . & . & . & . \\
. & . & . & . & . & . & . & . & .  \\
. & . & . & . & . & . & . & . & . \\
 \bottomrule
\bottomrule
\end{tabular}}
 }
{\caption{{\bf The first few entries of the multi-wavelength data presented in this work.} The entire dataset can be found in machine-readable format in the supplementary file named ``allphot.txt''. The {\bf Time} column lists days in observer frame since MJD 59621.4458. All optical/UV photometry ({\bf Flux} in milliJansky) has been corrected for MilkyWay extinction. \target's host galaxy was not detected in the pre-explosion panSTARRs images so host-subtraction was not performed. {\bf Observatory} is the name of the facility. Values of 1 and 0 in the ``Detection'' column indicate flux measurements and $3\sigma$ upper limits, respectively. }\label{tab: allphot}}

%%%%%%%%%%%%%%%%%%%%%%%%%%%%%%%%%%%%%%%%%%%%%%%%%%%
\ttabbox[\linewidth]{
\resizebox{\textwidth}{!}{\begin{tabular}{*{10}{l}}
\toprule
\toprule
\multicolumn{10}{c}{Best-fit parameters from fitting time-resolved 0.3-5.0 keV \nicer~ X-ray spectra} \\
\bottomrule
{\bf Start} & {\bf End} & {\bf Exposure} & {\bf FPMs} & {\bf Phase} & {\bf Index} & {\bf Log(Integ. Lum.)} & {\bf Log(Obs. Lum.)} & {\bf Count rate } & {\bf $\chi^{2}$/bins } \\
 (MJD) & (MJD) & (ks) & & & & (0.3-10 keV) & (0.3-5.0 keV) & (0.3-5.0 keV) & \\
\midrule
59626.75 & 59627.25 & 6.36 & 52 & E0 & 1.5$^{+0.01}_{-0.01}$ & 47.825$^{+0.003}_{-0.003}$ & 47.247$^{+0.003}_{-0.002}$ & 0.2354$\pm$0.0011 & 68.3/77 \\
59627.25 & 59627.75 & 5.28 & 52 & E1 & 1.58$^{+0.01}_{-0.01}$ & 47.715$^{+0.004}_{-0.004}$ & 47.099$^{+0.002}_{-0.004}$ & 0.1733$\pm$0.0011 & 97.4/73 \\
59627.75 & 59628.25 & 4.8 & 52 & E2 & 1.66$^{+0.01}_{-0.01}$ & 47.484$^{+0.005}_{-0.005}$ & 46.832$^{+0.002}_{-0.004}$ & 0.0971$\pm$0.001 &  112.6/72\\
59628.25 & 59628.75 & 5.76 & 52 & E3 & 1.65$^{+0.01}_{-0.01}$ & 47.613$^{+0.004}_{-0.004}$ & 46.965$^{+0.004}_{-0.002}$ & 0.1309$\pm$0.001 & 70.0/73 \\
59628.75 & 59629.25 & 3.48 & 52 & E4 & 1.64$^{+0.01}_{-0.01}$ & 47.496$^{+0.006}_{-0.006}$ & 46.851$^{+0.004}_{-0.004}$ & 0.1008$\pm$0.0013 & 83.7/71 \\
59629.25 & 59629.75 & 2.28 & 52 & E5 & 1.63$^{+0.02}_{-0.02}$ & 47.39$^{+0.008}_{-0.008}$ & 46.751$^{+0.006}_{-0.005}$ & 0.0801$\pm$0.0019 & 58.3/66 \\
59629.75 & 59630.25 & 2.64 & 52 & E6 & 1.69$^{+0.02}_{-0.02}$ & 47.405$^{+0.008}_{-0.008}$ & 46.737$^{+0.006}_{-0.004}$ & 0.0792$\pm$0.0018 &  70.4/67\\
59630.25 & 59630.75 & 2.76 & 51 & E7 & 1.69$^{+0.02}_{-0.02}$ & 47.483$^{+0.007}_{-0.007}$ & 46.818$^{+0.005}_{-0.004}$ & 0.0954$\pm$0.0017 & 64.2/69 \\
59630.75 & 59631.25 & 3.84 & 52 & E8 & 1.64$^{+0.01}_{-0.01}$ & 47.427$^{+0.006}_{-0.006}$ & 46.786$^{+0.004}_{-0.006}$ & 0.0865$\pm$0.0014 & 63.0/71 \\
59631.25 & 59631.75 & 5.64 & 52 & E9 & 1.61$^{+0.01}_{-0.01}$ & 47.377$^{+0.005}_{-0.005}$ & 46.747$^{+0.004}_{-0.003}$ & 0.0785$\pm$0.0009 & 86.8/72 \\
59631.75 & 59632.25 & 2.76 & 52 & E10 & 1.65$^{+0.02}_{-0.02}$ & 47.397$^{+0.007}_{-0.007}$ & 46.748$^{+0.004}_{-0.004}$ & 0.0801$\pm$0.0017 & 69.5/68\\
59632.25 & 59632.75 & 3.72 & 52 & E11 & 1.54$^{+0.02}_{-0.02}$ & 47.436$^{+0.007}_{-0.007}$ & 46.836$^{+0.005}_{-0.006}$ & 0.0696$\pm$0.0012 & 73.1/71\\
59632.75 & 59633.25 & 3.36 & 52 & E12 & 1.56$^{+0.02}_{-0.02}$ & 47.261$^{+0.007}_{-0.007}$ & 46.654$^{+0.005}_{-0.006}$ & 0.0621$\pm$0.0014 & 66.2/68\\
59633.25 & 59633.75 & 3.12 & 52 & E13 & 1.52$^{+0.02}_{-0.02}$ & 47.247$^{+0.007}_{-0.007}$ & 46.658$^{+0.005}_{-0.005}$ & 0.0617$\pm$0.0014 & 74.5/68\\
59633.75 & 59634.25 & 6.36 & 52 & E14 & 1.48$^{+0.01}_{-0.01}$ & 47.253$^{+0.005}_{-0.005}$ & 46.684$^{+0.003}_{-0.003}$ & 0.0643$\pm$0.0008 & 71.4/72\\
59634.25 & 59634.75 & 4.44 & 52 & E15 & 1.52$^{+0.02}_{-0.02}$ & 47.136$^{+0.007}_{-0.007}$ & 46.55$^{+0.007}_{-0.006}$ & 0.048$\pm$0.001 & 79.7/69\\
59634.75 & 59635.25 & 2.28 & 52 & E16 & 1.54$^{+0.02}_{-0.02}$ & 47.21$^{+0.009}_{-0.009}$ & 46.614$^{+0.006}_{-0.007}$ & 0.056$\pm$0.0019 & 62.5/63\\
59635.25 & 59635.75 & 1.8 & 52 & E17 & 1.55$^{+0.03}_{-0.03}$ & 47.128$^{+0.01}_{-0.011}$ & 46.529$^{+0.008}_{-0.008}$ & 0.0463$\pm$0.0024 & 50.6/58\\
59635.75 & 59636.25 & 2.16 & 52 & E18 & 1.54$^{+0.03}_{-0.03}$ & 47.009$^{+0.011}_{-0.011}$ & 46.414$^{+0.008}_{-0.011}$ & 0.0355$\pm$0.002 & 45.3/58\\
59636.25 & 59636.75 & 1.2 & 52 & E19 & 1.87$^{+0.05}_{-0.05}$ & 46.992$^{+0.02}_{-0.02}$ & 46.24$^{+0.013}_{-0.013}$ & 0.0272$\pm$0.0033 & 32.4/40\\
59636.75 & 59637.25 & 2.52 & 52 & E20 & 1.73$^{+0.03}_{-0.03}$ & 47.001$^{+0.013}_{-0.013}$ & 46.315$^{+0.01}_{-0.007}$ & 0.0306$\pm$0.0016 & 50.2/54\\
59637.25 & 59637.75 & 2.28 & 52 & E21 & 1.31$^{+0.03}_{-0.03}$ & 46.934$^{+0.011}_{-0.011}$ & 46.436$^{+0.013}_{-0.01}$ & 0.0349$\pm$0.0018 & 125.5/62\\
59637.75 & 59638.25 & 0.84 & 52 & E22 & 1.53$^{+0.06}_{-0.05}$ & 46.912$^{+0.02}_{-0.02}$ & 46.319$^{+0.016}_{-0.015}$ & 0.0288$\pm$0.0053 & 34.9/39\\
59638.25 & 59638.75 & 1.44 & 49 & E23 & 1.59$^{+0.04}_{-0.04}$ & 46.982$^{+0.015}_{-0.015}$ & 46.361$^{+0.013}_{-0.008}$ & 0.0322$\pm$0.0029 & 33.5/47\\
59638.75 & 59639.25 & 2.88 & 52 & E24 & 1.61$^{+0.03}_{-0.03}$ & 46.946$^{+0.011}_{-0.011}$ & 46.317$^{+0.01}_{-0.006}$ & 0.0293$\pm$0.0015 & 64.2/60\\
59639.25 & 59639.75 & 2.4 & 49 & E25 & 1.53$^{+0.04}_{-0.04}$ & 46.886$^{+0.013}_{-0.013}$ & 46.295$^{+0.007}_{-0.01}$ & 0.0272$\pm$0.0017 & 58.0/56\\
59639.75 & 59640.25 & 3.12 & 52 & E26 & 1.57$^{+0.03}_{-0.03}$ & 46.921$^{+0.011}_{-0.011}$ & 46.31$^{+0.009}_{-0.006}$ & 0.0284$\pm$0.0013 & 66.2/59\\
59640.25 & 59640.75 & 2.76 & 52 & E27 & 1.53$^{+0.03}_{-0.03}$ & 46.999$^{+0.01}_{-0.01}$ & 46.405$^{+0.008}_{-0.01}$ & 0.0347$\pm$0.0015 & 48.6/59\\
59640.75 & 59641.25 & 2.64 & 49 & E28 & 1.57$^{+0.03}_{-0.03}$ & 46.927$^{+0.012}_{-0.012}$ & 46.316$^{+0.013}_{-0.009}$ & 0.0286$\pm$0.0014 & 42.5/56\\
59641.25 & 59641.75 & 3.0 & 52 & E29 & 1.54$^{+0.03}_{-0.03}$ & 46.861$^{+0.012}_{-0.012}$ & 46.263$^{+0.009}_{-0.012}$ & 0.0252$\pm$0.0012 & 63.7/56\\
59641.75 & 59642.25 & 4.44 & 52 & E30 & 1.52$^{+0.03}_{-0.03}$ & 46.765$^{+0.011}_{-0.011}$ & 46.177$^{+0.01}_{-0.007}$ & 0.0206$\pm$0.0009 & 66.0/61\\
59642.25 & 59642.75 & 0.24 & 52 & E31 & 1.51$^{+0.15}_{-0.16}$ & 46.747$^{+0.052}_{-0.053}$ & 46.166$^{+0.042}_{-0.035}$ & 0.0208$\pm$0.0175 & 11.8/12\\
59642.75 & 59643.25 & 2.4 & 48 & E32 & 1.47$^{+0.05}_{-0.05}$ & 46.752$^{+0.016}_{-0.016}$ & 46.187$^{+0.014}_{-0.011}$ & 0.021$\pm$0.0019 & 70.5/56\\
 \bottomrule
\bottomrule
\end{tabular}}
 }
{\caption{{\bf Summary of time-resolved X-ray energy spectral modeling of \target}. Here,  0.3-5.0 keV \nicer~ spectra are fit with {\it tbabs*ztbabs*zashift(clumin*pow)} model using {\it XSPEC}\cite{xspec}. {\bf Start} and {\bf End} represent the start and end times (in units of MJD) of the interval used to extract a combined \nicer spectrum. {\bf Exposure} is the accumulated exposure time during this time interval. {\bf FPMs:} The total number of active detectors minus the ``hot'' detectors. {\bf Phase} is the name used to identify the epoch. {\bf Index} is the photon index of the power law component. {\bf Log(Integ. Lum.)} is the logarithm of the integrated absorption-corrected power law luminosity in 0.3-10 keV in units of erg s$^{-1}$. {\bf Log(Obs. Lum.)} is the logarithm of the observed 0.3-5.0 keV luminosity in units of erg s$^{-1}$. {\bf Count Rate } is the background-subtracted \nicer~ count rate in 0.3-5.0 keV in units of counts/sec/FPM. All errorbars represent 1-$\sigma$ uncertainties. {\bf $\chi^{2}$/bins} represents the best-fit $\chi^{2}$ and the number of spectral bins. The total $\chi^{2}$/degrees of freedom is 2135.3/1956.}\label{tab: xraydata}}

\ttabbox[\linewidth]{
\resizebox{\textwidth}{!}{\begin{tabular}{*{5}{c}}
\toprule
Model 1 & 59636.446 - 59638.446 & 59636.446 - 59638.446 & 59662.446 - 59667.446 & Tied \\
\midrule
B (G) & $0.13^{+0.03}_{-0.03}$ & $1.0^{+0.2}_{*}\times10^{-2}$ &  $9.7^{+5.4}_{-3.5}\times10^{-2}$ & \\
R (cm) & $5.9^{+0.2}_{-0.1}\times10^{15}$ & $6.9^{+0.3}_{-0.3}\times10^{15}$ & $1.0^{*}_{-0.3}\times10^{16}$ & \\ 
$\rm{n_e}$ (cm$^{-3}$) & $973^{+195}_{-160}$ & $2200^{+237}_{-205}$ &  $144^{+58}_{-38}$ & \\
$\gamma_{\rm max}$ & $5.0^{+1.2}_{-0.9}\times10^{3}$ & $3.2^{+1.8}_{-0.4}\times10^{4}$ & $3.4^{+1.4}_{-0.9}\times10^{3}$ & \\ 
$\gamma_{\rm min}$ & & & & $91^{+4}_{-4}$ \\ 
$p$ & & & & $2.21^{+0.05}_{-0.05}$ \\
$\Gamma_{\rm j}$ & & & & $86^{+9}_{-10}$ \\ 
$\theta$ & & & & $0.5^{+0.1}_{*}$ \\
$\rm{Lum_{bb}}$ (erg/s) & & & &  $1.71^{+0.13}_{-0.11}\times10^{45}$ \\ 
$T_{\rm bb}$ (K) & & & &  $2.34^{+0.16}_{-0.14}\times10^{4}$\\ 
\midrule 
$\delta $ & & & & $103$\\
$\rm{P_e}$ (erg/s) & $5.3\times10^{45}$ & $2.0\times10^{46}$ & $2.0\times10^{45}$ & \\
$\rm{P_b}$ (erg/s) & $1.6\times10^{43}$ & $1.5\times10^{41}$ & $2.6\times10^{43}$ & \\
$\rm{P_p}$ (erg/s) & $3.6\times10^{46}$ & $1.1\times10^{47}$ & $1.5\times10^{46}$ & \\
$\rm{P_j}$ (erg/s) & $4.1\times10^{46}$ & $1.3\times10^{47}$ & $1.7\times10^{46}$ & \\
$\rm{U_e/U_b}$ & $325$ & $1.3\times10^{5}$ & $77$ & \\
$\rm{R_{bb}}$ (cm) & &  &  & $2.8\times10^{15}$\\
\midrule
Model 2 & 59636.446 - 59638.446 & 59636.446 - 59638.446 & 59662.446 - 59667.446 & Tied \\
\midrule
B (G) & $10.2^{+2.0}_{-1.6}$ & $18^{+5}_{-3}$ &  $36^{+14}_{-9}$ & \\
R (cm) & $1.16^{+0.12}_{-0.10}\times10^{14}$ & $6.0^{+0.9}_{-0.8}\times10^{13}$ & $2.2^{+0.4}_{-0.6}\times10^{14}$ & \\ 
$\rm{n_e}$ (cm$^{-3}$) & $8.7^{+1.5}_{-1.3}\times10^{7}$ & $1.3^{+0.3}_{-0.3}\times10^{8}$ &  $4.2^{+2.0}_{-1.5}\times10^{6}$ & \\ 
$\gamma_{\rm max}$ & $1.2^{+0.9}_{-0.4}\times10^{4}$ & $3.4^{+2.2}_{-1.3}\times10^{3}$ & $6.7^{+2.3}_{-1.7}\times10^{2}$ & \\ 
$\gamma_{\rm min}$ & & & & $4.7^{+0.5}_{-0.4}$ \\ 
$p$ & & & & $2.13^{+0.09}_{-0.08}$ \\
$\Gamma_{\rm j}$ & & & & $5^{+1}_{-*}$ \\ 
$\theta$ & & & & $1.3^{+0.8}_{-0.6}$ \\
$\rm{Lum_{bb}}$ (erg/s) & & & &  $1.36^{+0.10}_{-0.08}\times10^{45}$ \\ 
$T_{\rm bb}$ (K) & & & &  $2.10^{+0.11}_{-0.10}\times10^{4}$\\ 
\midrule 
$\delta $ & & & & $10.7$\\
$\rm{P_e}$ (erg/s) & $4.5\times10^{43}$ & $2.3\times10^{43}$ & $7.6\times10^{42}$ & \\
$\rm{P_b}$ (erg/s) & $1.6\times10^{41}$ & $1.4\times10^{41}$ & $6.9\times10^{42}$ & \\
$\rm{P_p}$ (erg/s) & $5.0\times10^{45}$ & $2.0\times10^{45}$ & $8.2\times10^{44}$ & \\
$\rm{P_j}$ (erg/s) & $5.1\times10^{45}$ & $2.0\times10^{45}$ & $8.4\times10^{44}$ & \\
$\rm{U_e/U_b}$ & $412$ & $164$ & $1.1$ & \\
$\rm{R_{bb}}$ (cm) & &  &  & $3.1\times10^{15}$\\
\bottomrule
\end{tabular}}
 }
{\caption{{\bf Summary of the best-fitting jet models}. The emitting region magnetic field $B$, radius $R$ and number density $\rm{n_e}$, as well as the maximum Lorentz factor of the particles $\gamma_{\rm max}$ were left free to vary in each epoch. The minimum electron Lorentz factor $\gamma_{\rm min}$, particle distribution slope $p$, jet bulk Lorentz factor $\Gamma_{\rm j}$, viewing angle $\theta$, black body luminosity $L_{\rm bb}$ and black body temperature $T_{\rm bb}$ were tied. The parameters marked with a $*$ were pegged to their limit. The statistic for the overall joint fit is $\chi^{2}/\text{d.o.f.}=305.54/138=2.20$ for model 1 and $284.45/123=2.31$ for model 2. We also report the power carried by the electrons, protons (assuming one cold proton per electron) and magnetic field $P_e$, $P_p$, $P_b$, the total jet power $P_j=P_e+P_p+P_b$, the equipartition fraction $U_e/U_b$, and the black body radius $R_{\rm bb}$.}\label{tab: SED_fits}}

%%%%%%%%%%%%%%%%%%%%%%%%%%%%%%%%%%%%%%%%%%%%%%%%%%%%%%%%%%%%%%%%%%%%%

%%%%%%%%%%%%%%%%%%%%%%%%%%%%%%%%%%%%%%%%%%%%%%%%%%%%%%%%%%%%%%%%%%%%%%

% \section*{{\Huge Supplementary Methods.}}

\end{document}